\documentclass{jcgt}

\usepackage{stix2}
\usepackage{hyperref}
\usepackage{cancel}
\usepackage{gensymb}
\usepackage{upgreek}

\setciteauthor{Purva Kulkarni, Aravind Shankara Narayanan}
\setcitetitle{A Comprehensive Guide to Mesh Simplification using Edge Collapse}

\submitted{\today}

\begin{document}

\title{A Comprehensive Guide to Mesh Simplification using Edge Collapse }

\author
       {Purva Kulkarni~\href{https://orcid.org/0009-0006-3849-137X}{\includegraphics[width=8pt]{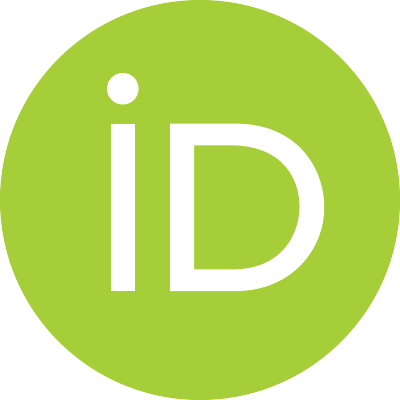}}\\Independent Researcher
        \and Aravind Shankara Narayanan ~\href{https://orcid.org/0000-0002-2472-7402}{\includegraphics[width=8pt]{ORCIDlogo}}\\Independent Researcher
       }

\teaser{
  \includegraphics[width=0.75\textwidth]{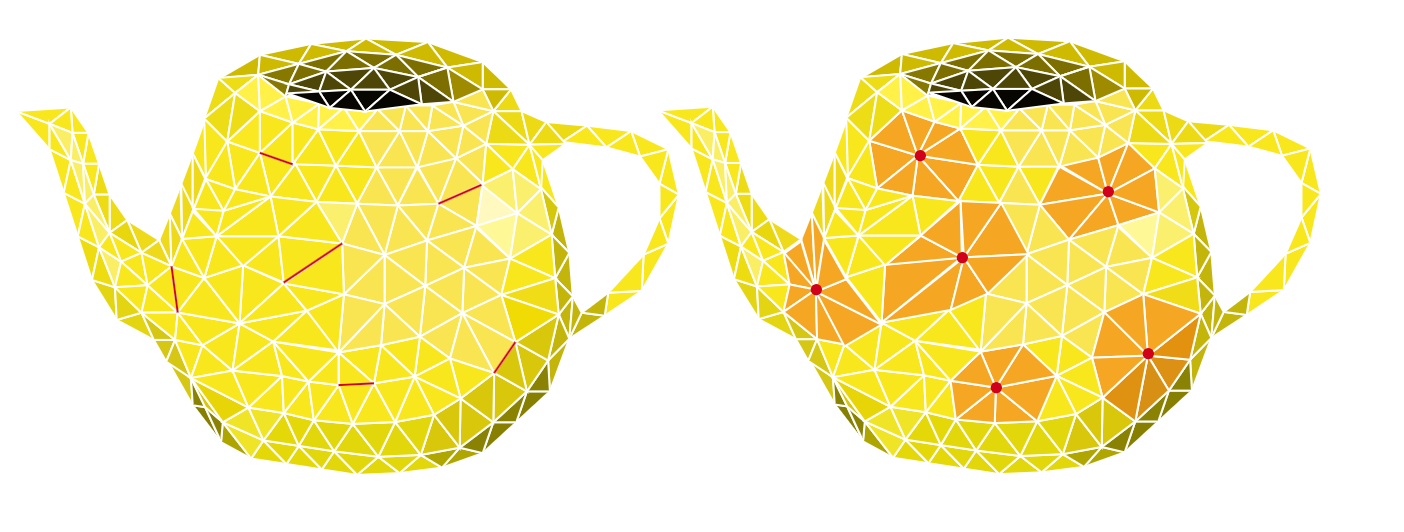}
  \caption{Mesh simplification using edge collapse.}
  \label{fig:teaser}
}

\maketitle

\begin{abstract}
\textit{{\small Mesh simplification is the process of reducing the number of vertices, edges and triangles in a three-dimensional (3D) mesh while preserving the overall shape and salient features of the mesh. A popular strategy for this is edge collapse, where an edge connecting two vertices is merged into a single vertex. The edge to collapse is chosen based on a cost function that estimates the error introduced by this collapse. This paper presents a comprehensive, implementation-oriented guide to edge collapse for practitioners and researchers seeking both theoretical grounding and practical insight. We review and derive the underlying mathematics and provide reference implementations for foundational cost functions including Quadric Error Metrics (QEM) and Lindstrom-Turk’s geometric criteria. We also explain the mathematics behind attribute-aware edge collapse in QEM variants and Hoppe’s energy-based method used in progressive meshes. In addition to cost functions, we outline the complete edge collapse algorithm, including the specific sequence of operations and the data structures that are commonly used. To create a robust system, we also cover the necessary programmatic safeguards that prevent issues like mesh degeneracies, inverted normals, and improper handling of boundary conditions. The goal of this work is not only to consolidate established methods but also to bridge the gap between theory and practice, offering a clear, step-by-step guide for implementing mesh simplification pipelines based on edge collapse.}}
\end{abstract}

\section{Introduction}

{\small Triangles are the most commonly used drawing primitive in computer graphics. They are natively supported by almost all graphics libraries and hardware systems, making triangular meshes the dominant representation in 3D modeling. Modern graphics systems are capable of rendering models composed of millions of triangles, thanks to decades of hardware advancements. However, with Moore’s Law plateauing and the geometric complexity of meshes increasing rapidly, relying on brute-force parallel processing is no longer viable. This makes mesh simplification techniques more essential than ever for achieving real-time performance and scalability in interactive and large-scale applications. Mesh simplification forms the basis of level of detail (LOD) systems to ease GPU workload, accelerates collision detection in games, and enables faster coarse approximations in FEA simulations.}

{\small Among the various mesh simplification techniques available, edge collapse is most widely adopted in practice. This strategy is implemented in many major graphics libraries and tools like \textit{CGAL} , \textit{QSlim}, and \textit{meshoptimizer}. An edge collapse operation merges the two endpoints of an edge into a single new vertex, effectively removing the edge and the two triangles that shared it. Repeating this operation iteratively leads to a simplified mesh that maintains the overall structure of the original. Cost functions help determine which edge to collapse and where to place the resulting vertex in order to best preserve the model’s visual and geometric details.}

\begin{figure}[!ht]
\centering
   \includegraphics[width=0.65\columnwidth]{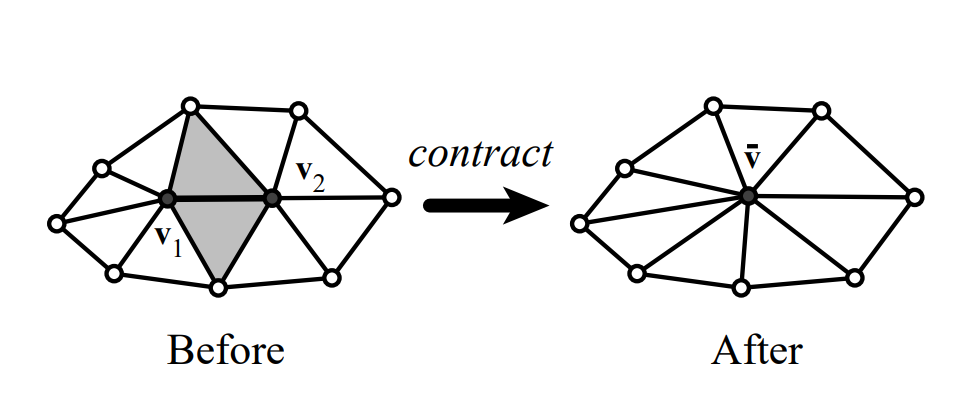}
   \caption{\label{fig:edge-collapse} Edge collapse}
\end{figure}

{\small While mesh simplification is a well-studied topic, newcomers to the field often face a steep learning curve when engaging with foundational papers. Many of these works emphasize final equations or high-level algorithmic descriptions, offering little insight into the underlying geometric reasoning or the practical implementation details. As a result, readers may struggle to build an intuitive understanding of how and why edge collapse-based simplification works, or how to translate theory into working code.}

{\small This paper aims to bridge that gap by offering a detailed, implementation-aware analysis of edge collapse-based mesh simplification on a manifold mesh. Our contributions are as follows:}

\begin{itemize}
\item {\small We present a complete, end-to-end simplification pipeline that includes well-chosen data structures for representing mesh connectivity, deep analysis of cost functions presented in foundational papers in this space, and the edge collapse algorithm that binds both of these.}
\item {\small Unlike many prior works that present only the final cost metrics or optimization functions, we derive and explain them along with the geometric meaning behind these formulations, allowing readers to understand the rationale behind each step.}
\item {\small Our goal is two-fold: to serve as a conceptual guide for learners who want to understand the inner workings of simplification algorithms, and to act as a practical reference for developers looking to implement their own systems.}
\end{itemize}

{\small In this paper, we first categorize and review different families of mesh simplification algorithms. Since the efficiency of edge-collapse operations depends on fast access to mesh connectivity and rapid local updates, int the following section, we discusse data structures that can be employed to store and manage the mesh connectivity information. We then present a comprehensive edge-collapse algorithm, including detailed programmatic checks to prevent mesh degeneracies. Our most extensive section examines cost computation strategies, explaining the mathematical formulations from foundational papers alongside practical implementations. In the next section, we cover advance edge collapse techinques that account for per-vertex attributes. Finally, we provide supplemental mathematical results and proofs that support these techniques.}

\section{Families of mesh simplification techniques }

{\small Mesh simplification techniques vary widely, but most can be grouped by the strategy they use to reduce geometric complexity while maintaining topology as presented in \cite{cignoni1998comparison}. We have supplemented this list with recent advances in the field that leverage modern techniques such as machine learning and neural networks.}

{\small An early strategy for mesh decimation focused on detecting coplanar or nearly coplanar surface patches and merging them into larger polygonal regions as presented in \cite{dehaemer1991simplification} and \cite{hinker1993geometric}. These regions are subsequently re-triangulated to produce a mesh with fewer faces. Despite its simplicity, the method often degraded geometric detail and introduced topological inconsistencies.}

{\small Another method, known as vertex clustering, groups nearby vertices based on spatial proximity and replaces each cluster with a single representative vertex, followed by local re-triangulation as presented in \cite{rossignac1993multi} and improved in \cite{low1997model}. While faster, this method was again found to compromise detail and topological accuracy.}

{\small A more refined and topology-sensitive method is iterative local decimation, which incrementally removes vertices, edges, or faces based on localized geometric evaluations. These operations are typically guided by cost functions designed to preserve the mesh’s overall structure and appearance \cite{Garland1997Surface, Lindstrom1998Turk, schroeder1992decimation}. Extensions such as simplification envelopes \cite{cohen1996simplification} presents bounded error control by forcing the resulting simplified mesh to lie between two offset meshes.}

{\small In energy-based optimization methods, such as the one presented in \cite{hoppe1993mesh}, a global cost function evaluates the overall quality of the mesh. Simplification is carried out through iterative edge-based operations such as collapse, swap, or split that aims at minimizing both the local and global cost function. Although this approach with global optimization promises a better overall structural preservation, it is less commonly used in practice due to its computational complexity.}

{\small A different strategy is retiling, introduced in \cite{turk1992re}, which begins by randomly placing a user-defined reduced number of new vertices on the original surface which are then adjusted based on areas of high curvature. A new reduced triangulation is built on this vertex set. Although effective in reducing triangle count, this method lacks support for per-vertex attributes, making it less suitable for applications like computer-aided design (CAD) or physical simulations where such data is essential.}

{\small Another notable approach to mesh simplification is voxelization, as used by works such as \cite{he1995voxel} and \cite{he1996controlled}. Here, the mesh is first sampled into a voxel grid, and a low-pass filter is applied at each grid point to generate a discrete scalar field. A triangulated surface is then extracted from this field using the standard marching cubes algorithm or an adaptive variant of it at an isovalue dictated by the filter. The detail of voxel-based meshes can be adjusted via resolution, but the method sees limited industrial use. It smooths sharp features making itself unsuitable for CAD and is computationally expensive due to volumetric processing, and lacks explicit geometric error control, making output quality difficult to guarantee.}

{\small Recent work has explored neural methods that either simplify meshes directly or offer implicit representations enabling level-of-detail control. \cite{potamias2022neural} employs a differentiable neural network to select a subset of input vertices using a sparse attention mechanism and re-triangulate the selected vertices, producing simplified meshes in a data-driven, generalizable manner without per-mesh retraining. \cite{chen2023neural} generates a coarse base mesh using QEM, followed by neural remeshing through face splits. A per-face latent feature representation is transmitted and decoded on the client-side to reconstruct finer meshes. This approach implicitly generates simplified representations across multiple LODs. \cite{park2019deepsdf} learns a signed distance field (SDF) representation from a voxelized representation of mesh. \cite{takikawa2021neural} extends it by creating multiscale SDFs giving real-time rendering at various LODs via ray marching. Although simplified triangle meshes can be extracted using methods like marching cubes, this undermines the efficiency of its implicit representation.}

\section{Data structures representing mesh connectivity}

{\small Mesh connectivity data structures are designed to efficiently organize and manage the relationships between elements of a mesh such as which faces share an edge, which edges are connected to a vertex, or which vertices make up a face. They allow algorithms to rapidly traverse and manipulate the mesh’s topology. Below are two data structures commonly used to represent mesh connectivity, along with an evaluation of their suitability for supporting edge collapse operations.}
\newline
{\small The \textit{Corner Table} data structure introduced in \cite{rossignac2002edgebreaker} is a compact mesh representation where each triangle's three "corners" (vertex-triangle associations) are stored in a list. For edge collapse, it efficiently manages the edge-collapse updates and supports fast querying on the mesh.}
\newline
{\small The \textit{Half-Edge} data structure presented in \cite{mcguire2000half}, is widely used due to its intuitive design and broad support across mesh libraries. In this structure, each mesh edge is represented by a pair of half-edges pointing in opposite directions, each storing connectivity to associated elements such as vertices, faces, and neighboring edges. While not the most memory-efficient option, it enables fast mesh queries and local updates, making it ideal for operations like edge collapse. }
\newline
{\small In the code listing below, we present the interfaces that a typical connectivity data structure would support.}
\newline
\begin{lstlisting}
class IVertex 
{
  virtual vec3 GetPosition() = 0;
  virtual vecn GetAttributes() = 0;
};

class IEdge 
{
  virtual vector<IVertex*> GetVertices() = 0;
};

class IFace 
{
  virtual vector<IVertex*> GetVertices() = 0;
  virtual vector<IEdge*> GetEdges() = 0;
};

class IMesh
{
  virtual vector<Face*> GetConnectedFaces(Edge* edge) = 0;
  virtual vector<Face*> GetConnectedFaces(Vertex* vertex) = 0;
  virtual vector<Edge*> GetConnectedEdges(Vertex* vertex) = 0;
  virtual vector<Vertex*> GetConnectedVertices(Vertex* vertex) = 0;
};
\end{lstlisting}

{\small The queries listed in table. \ref{tab:mesh-query-operations} are necessary for the edge collapse-based mesh simplification algorithm, so they must be handled efficiently by the chosen mesh connectivity data structure. As \autoref{tab:mesh-query-operations} illustrates, both the Half-Edge and Corner Table structures are adept at handling these queries with optimal time complexities, making them well-suited for edge-collapse based mesh simplification.}

\begin{center}
\begin{table}[!h]
\begin{tabular}{|c|c|} 
\hline
\textbf{Mesh Query} & \textbf{Time complexity} (half-edge / corner table) 
\\ 
\hline
Get all triangles connected to vertex $v$ & $O(\text{degree}(v))$ 
\\ 
\hline
Get all edges connected to vertex $v$ & $O(\text{degree}(v))$
\\
\hline
Get all vertices connected to vertex $v$ & $O(\text{degree}(v))$
\\
\hline
Get all edges connected to vertex $v$ & $O(\text{degree}(v))$
\\
\hline
\end{tabular}
\caption{\label{tab:mesh-query-operations} Mesh query operations and their time complexities using different data structures}
\end{table}
\end{center}

\section{Edge collapse-based simplification algorithm}

{\small Edge collapse-based simplification iteratively reduces the number of triangles in a mesh while preserving its overall shape and features. The core algorithm remains largely consistent across different implementations, with key differences lying in the cost metric and vertex placement strategies. The algorithm typically involves the following steps:}
\newline
\newline
\noindent \textbf{\small{Cost assignment and optimal vertex placement calculation}}

\begin{enumerate}
\item {\small A cost is computed for each edge in the mesh to estimate the geometric error introduced by collapsing it. Simultaneously, the optimal position for the resulting merged vertex is determined. This step is critical, as it is where most edge collapse based simplification strategies diverge. }
\item {\small The computed cost, along with the edge and its optimal replacement vertex, is stored in a priority queue.}
\end{enumerate}

\noindent \textbf{{\small Iterative edge collapse}}

{\small A target triangle count is either defined internally by the program or specified externally by the client code. Then, the following steps are repeated until the target triangle count is reached:}

\begin{enumerate}
\item {\small Select the edge with the lowest collapse cost from the priority queue.}
\item {\small Perform validity checks to ensure that collapsing that edge preserves the mesh’s manifoldness. (The three validity checks we employ are explained below.)}
\item {\small If the edge passes all validity checks, collapse it by replacing the edge with the computed vertex and removing the two adjacent triangles.}
\item {\small Since the collapse locally alters the mesh, recompute the costs of all the edges connected to the collapsed edge, and update the corresponding entries in the priority queue to maintain accuracy for the next iteration.}
\item {\small Update the mesh’s connectivity data structure to reflect the changes made by the collapse.}
\end{enumerate}

\noindent\textbf{{\small Edge collapse validity checks}}

{\small Checks 1 and 2 follow the criteria established in \cite{hoppe1993mesh}, while check 3 is derived empirically. These checks are crucial for avoiding degeneracies that may result in invalid or non-manifold mesh structures.}

\begin{enumerate}

\begin{figure}[!ht]
\centering
   \includegraphics[width=0.65\columnwidth]{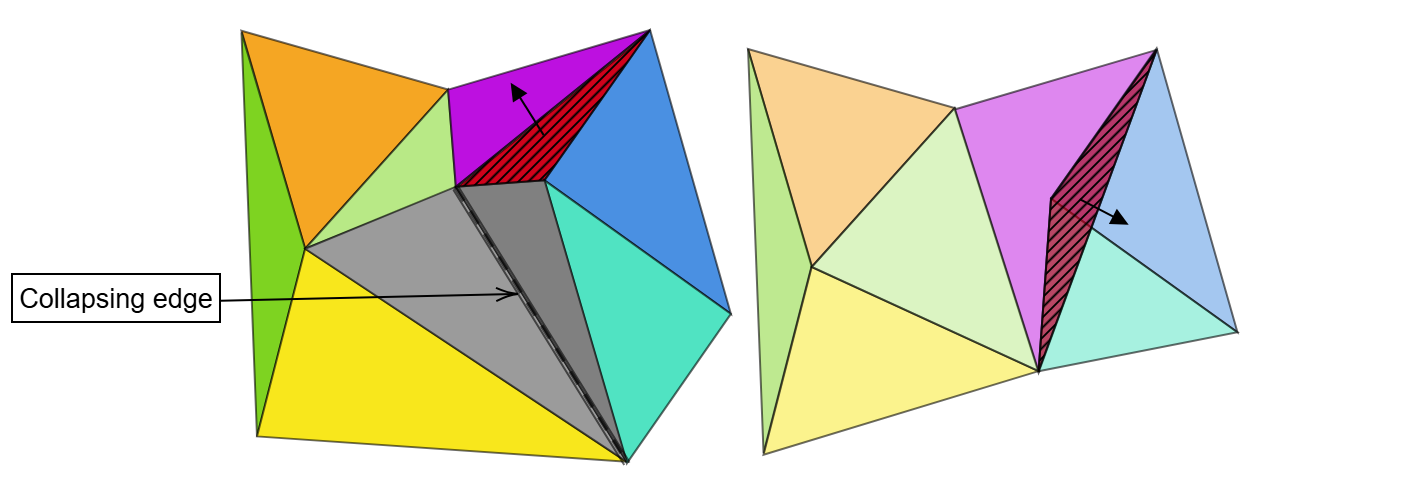}
   \caption{\label{fig:flipped-red-triangle} Flipped red triangle, causing local mesh degeneracy}
\end{figure}

\item {\small \textbf{Triangle flip check}: Ensure that collapsing the edge does not invert the orientation of any surrounding triangles. An edge collapse that results in such an inversion is shown in \autoref{fig:flipped-red-triangle}.}

\begin{lstlisting}
bool AreFacesFlipped(const IFace* old_face, const IFace* new_face)
{
  return dot(util::ComputeNormal(old_face), util::ComputeNormal(new_face)) < 0.0;
}
\end{lstlisting}

\begin{figure}[!ht]
\centering
   \includegraphics[width=0.5\columnwidth]{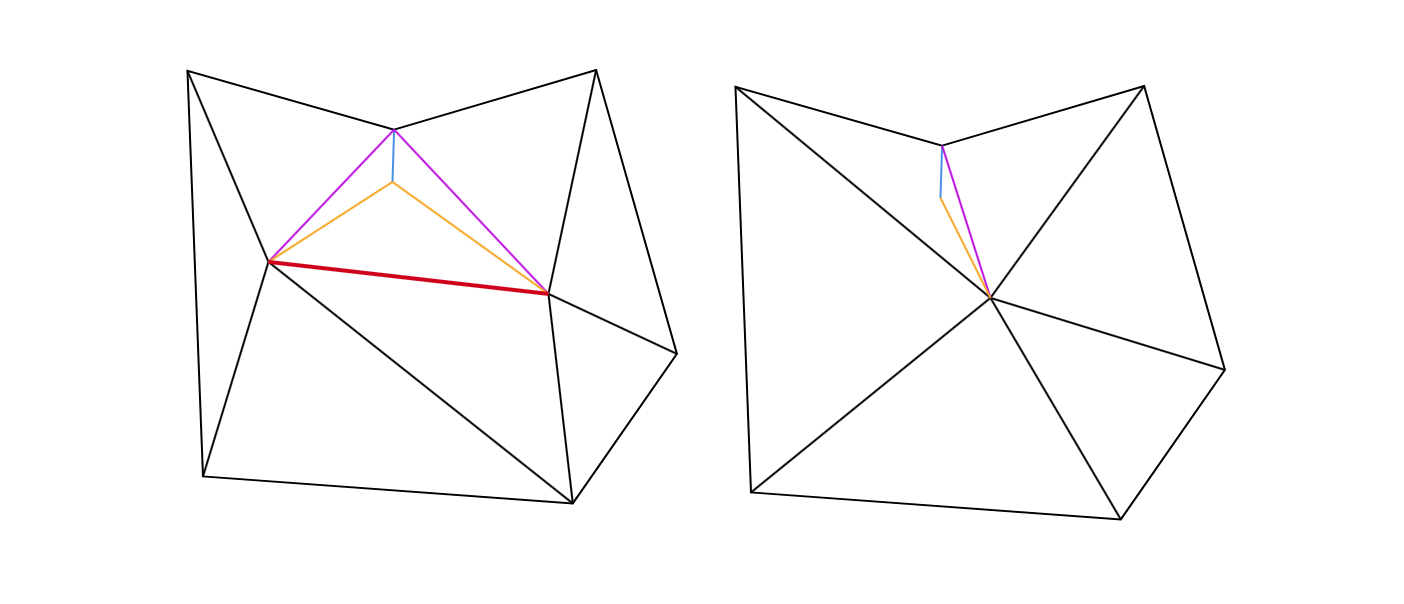}
   \caption{\label{fig:non-manifold-formation} Non-manifold triangle formation caused by collapsing the red edge}
\end{figure}

\item {\small \textbf{Two-neighbor connectivity check}: Verify that exactly one pair of edges is merged on each side of the collapsing edge. This condition holds when the two collapsing vertices share exactly two common neighbors. A connectivity-related non-manifold triangle formation is illustrated in  \autoref{fig:non-manifold-formation}.}

\begin{lstlisting}
bool HasMoreThanTwoNeighbors(const IMesh* mesh, const IEdge* collapse_edge)
{
  auto vertices = collapse_edge->GetVertices();
  auto conn_verts_v0 = mesh->GetConnectedVertices(vertices[0]);
  auto conn_verts_v1 = mesh->GetConnectedVertices(vertices[1]);
  
  set<IVertex*> conn_verts_v1_set(conn_verts_v1.begin(), conn_verts_v1.end());  

  int common_count = 0;
  for(auto vert : conn_verts_v0)
    if(conn_verts_v1_set.contains(vert))
    {
      common_count++;
      if (common_count > 2) return true; 
    }
  return false;
}
\end{lstlisting}

\begin{figure}[!ht]
\centering
   \includegraphics[width=0.5\columnwidth]{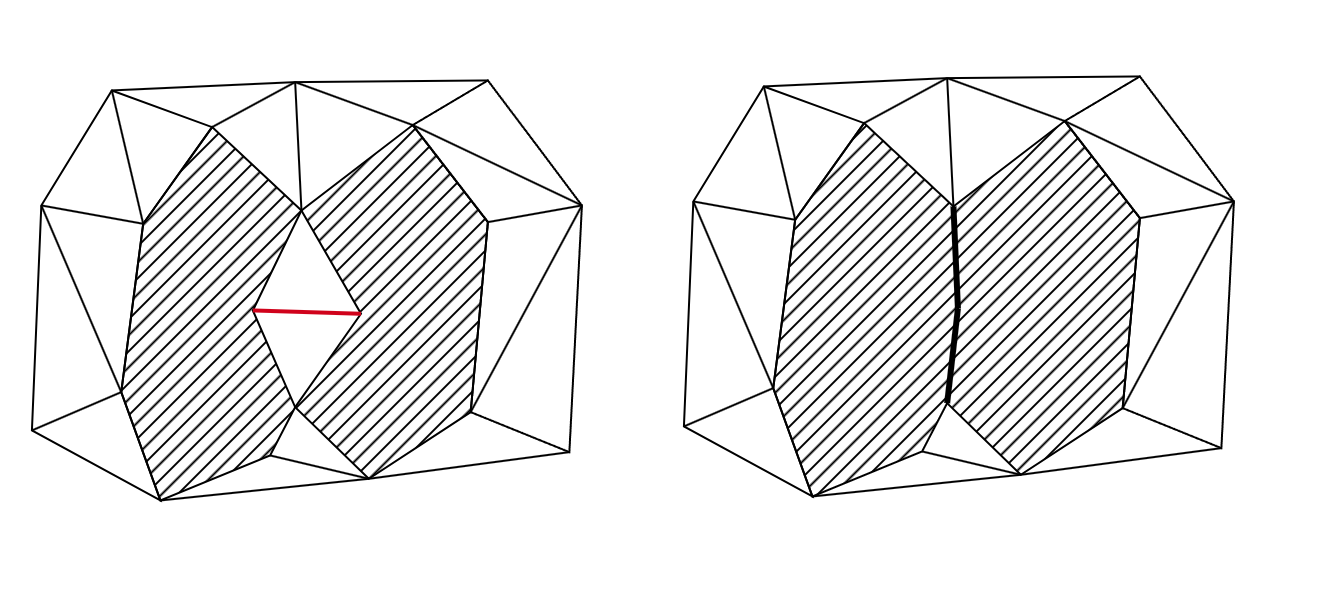}
   \caption{\label{fig: merged-boundaries-check} Edge collapse merging internal boundaries forming ill-formed local mesh}
\end{figure}

\item {\small \textbf{Boundary merge check}: If both the other edges of a face besides the collapsing edge lie on mesh boundaries or holes, collapsing the edge can lead to merging of two boundaries causing an ill-formed structure as illustrated in \autoref{fig: merged-boundaries-check}.}

\begin{lstlisting}
bool HasMultipleConnectedBoundaries(const IMesh* mesh, const IEdge* collapse_edge)
{
  for(auto face : mesh->GetConnectedFaces(collapse_edge))
  {
    int n_edges_on_boundary = 0;
    for(auto edge : face->GetEdges())
    {
      if(edge == collapse_edge) continue;
      if(util::IsBoundaryEdge(edge)) n_edges_on_boundary++;
    }
    if(n_edges_on_boundary > 1) return true;
  }
  return false;
}
\end{lstlisting}

\end{enumerate}

\section{Cost functions}

{\small In edge collapse-based mesh simplification, an \textit{error metric} is assigned to each edge that estimates the cost of collapsing it. Edges with the lowest error are prioritized for collapse. Additionally, we need effective strategies to determine the \textit{best new vertex position} that will replace the collapsed edge while minimizing the geometric distortion.}

{\small The {\fontfamily{pcr}\selectfont IConstraint} class below defines an interface for error metrics. The cost function classes implementing this interface compute the cost $\Eulerconst(v)$ of collapsing an edge for a candidate vertex position $v$. Implementations of this class compute the cost as $\Eulerconst(v)=v^THv+2c^Tv+k$, and store the entities $\{H,c,k\}$ in this equation in {\fontfamily{pcr}\selectfont m\_H, m\_c, m\_k}. These will be used to obtain the optimal vertex placement as well, as detailed in \autoref{sec:vertex-placement}.}

\begin{lstlisting}
class IConstraint
{
  public:
    virtual void EvaluateCost(const vec3 &v) = 0;

    const mat3& GetH() const { return m_H; }
    const vec3& GetC() const { return m_c; }

  private:
    mat3 m_H;
    vec3 m_c;
    double m_k;
}
\end{lstlisting}

\subsection{Plane-based quadric error metrics (QEM)}

{\small This method, described in \cite{Garland1997Surface}, defines error as the sum of distances from the new vertex to the planes of surrounding triangles, treating each vertex as their intersection. This captures how much the new vertex deviates from the original geometry, reflecting the introduced distortion.}

\begin{figure}[!ht]
\centering
   \includegraphics[width=0.65\columnwidth]{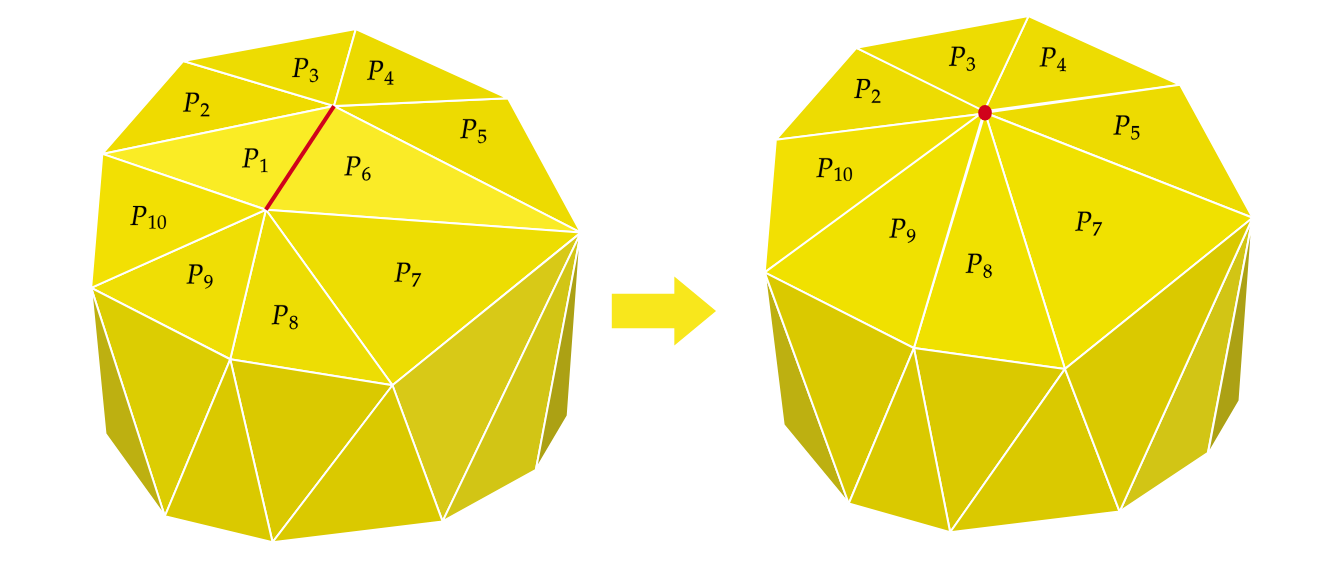}
   \caption{\label{fig:qem-local-structure} Local structure around edge collapse for QEM}
\end{figure}

{\small Referring to \autoref{fig:qem-local-structure}, we collapse the edge $(v_1,v_2)$ into a new vertex $v$, removing the two adjacent triangles(in planes $P_1$ and $P_6$) and forming a local geometric approximation. The error is measured as the sum of distances from $v$ to the original surrounding planes $P_1$ through $P_{10}$.}

{\small Let $\mathbb{P} =\{P_1 ,P_2 ,\dotsc ,P_{m}\}$, the set of planes that surround the edge being collapsed. Let $\Eulerconst $ be the error introduced by the newly added vertex $v$, given by:}

{\small \begin{equation*}
{\Eulerconst \ =\ \sum _{( n,d) \in \mathbb{P}}\left( n^{T} v+d\right)^{2}}
\end{equation*}}
{\small where $( n,d)$ represent the unit normal $n$ and scalar $d$ in the equation $n\cdotp r+d=0$ of each plane $P\in \mathbb{P}$.}

{\small The term ${n^{T} v+d}$ gives the signed distance from the vertex to the plane. Squaring it ensures that the error is always non-negative, penalizing both positive and negative deviations equally.}

{\small Expanding, 
\begin{equation*}
\begin{aligned}
{\Eulerconst } & ={\sum _{( n,d) \in \mathbb{P}}\left( v^{T} n+d\right)\left( n^{T} v+d\right)} & \\
 & {=\sum _{( n,d) \in \mathbb{P}}\left( v^{T}\left( nn^{T}\right) v+2dn^{T} v+d^{2}\right)} & \\
 & =v^{T} \textcolor[rgb]{0.82,0.01,0.11}{\left({\textstyle{\sum _{( n,d) \in \mathbb{P}}}{nn}{^{T}}}\right)} v + 2 \textcolor[rgb]{0.29,0.56,0.89}{\left({{\sum _{( n,d) \in \mathbb{P}}}{dn}}\right)}^T v + \textcolor[rgb]{0.74,0.06,0.88}{\textstyle{\sum _{(n,d) \in \mathbb{P}}{d^2}}} & \\
 & =v^{T}\textcolor[rgb]{0.82,0.01,0.11}{H} v+2\textcolor[rgb]{0.29,0.56,0.89}{c}^{T} v+\textcolor[rgb]{0.56,0.07,1}{k} 
\end{aligned}
\end{equation*}}

{\small To minimize the error, we set its gradient to zero and solve for $v$:}

{\small \begin{equation*}
\begin{aligned}
\nabla \Eulerconst  & =2Hv+2c=0\\
\Longrightarrow v & =-H^{-1} c
\end{aligned}
\end{equation*}}

\noindent {\small \textbf{Note:} When $\Eulerconst$ takes this form, the matrix $H$ is its Hessian matrix. In this specific case, $H$ turns out to be positive semidefinite. So, the point that makes $\nabla \Eulerconst=0$ corresponds to a minimum point rather than a maximum or a saddle point.}

{\small If the matrix $H$ is non-invertible (i.e., $\det( H) =0$), the optimal vertex position cannot be computed this way. In such cases, fallback strategies or alternative constraints are used. For this cost function, a non-invertible $H$ indicates that the surface surrounding the edge collapse is flat, as explained below:}

{\small \begin{equation*}
\begin{aligned}
H & ={\sum nn^{T}} & \text{where } n=\begin{pmatrix}
a & b & c
\end{pmatrix}^{T}\\
 & =\sum \begin{pmatrix}
a\\
b\\
c
\end{pmatrix}\begin{pmatrix}
a & b & c
\end{pmatrix} & \\
 & =\sum \begin{pmatrix}
a^{2} & ab & ac\\
ab & b^{2} & bc\\
ac & bc & c^{2}
\end{pmatrix} & 
\end{aligned}
\end{equation*}}

{\small Observe that $\begin{pmatrix}
a^{2} & ab & ac\\
ab & b^{2} & bc\\
ac & bc & c^{2}
\end{pmatrix} =\begin{pmatrix}
an & bn & cn
\end{pmatrix}$ in block notation. This means that each term of the sum is itself a non-invertible matrix, as all its columns are parallel to $n$. So, when $H$ is non-invertible, all the normals of the planes forming $H$ are parallel. This occurs if the local surface is flat.}\newline

\noindent {\small \textbf{Note:} The "quadric" in the name of this method is derived from the form this error takes when $v$ is represented in homogeneous 4-dimensional coordinates as the vector $\begin{pmatrix}
v\\
1
\end{pmatrix}$. In that case, the error $\Eulerconst $ is expressed as follows:}

{\small \begin{equation*}
\Eulerconst =\begin{pmatrix}
v & 1
\end{pmatrix}^{T}\begin{pmatrix}
H & c\\
c^{T} & k
\end{pmatrix}\begin{pmatrix}
v\\
1
\end{pmatrix}
\end{equation*}}

{\small where the authors define the 4x4 matrix $Q=\begin{pmatrix}
H & c\\
c^{T} & k
\end{pmatrix}$ above as the \textit{total error quadric} for this edge. It can further be decomposed as the sum of \textit{fundamental error quadrics} $K_{P}$ for each plane $P\in \mathbb{P}$:}

{\small \begin{equation*}
Q=\begin{pmatrix}
H & c\\
c^{T} & k
\end{pmatrix} ={\sum _{P\in \mathbb{P}} K_{P} ,\ K_{P} =\begin{pmatrix}
nn^{T} & dn^{T}\\
dn & d^{2}
\end{pmatrix}}
\end{equation*}}

{\small However, the same authors in \cite{Garland1998Simplifying} found this formulation impractical because it requires computationally expensive matrix operations on higher-dimensional matrices, like inversion. For this reason, it won't be discussed further.}

\begin{lstlisting}
class QEM : IConstraint
{
  void QEM(const IMesh* mesh, const IEdge* collapse_edge)
  {
    auto vertices = collapse_edge->GetVertices();
    auto connected_faces = util::GetUnion(
      mesh->GetConnectedFaces(vertices[0]),
      mesh->GetConnectedFaces(vertices[1])
    );

    m_H = mat3(0); m_c = vec3(0); m_k = 0;
    for (auto face : connected_faces) 
    {
      auto face_normal = util::ComputeNormal(face);
      auto v0 = face->GetVertices()[0];
      double d = -dot(face_normal, v0->GetPosition());

      m_H += outerProduct(face_normal, face_normal);
      m_c += d * face_normal;
      k += d * d;
    }
  }

  double EvaluateCost(const vec3& v) const override 
  { 
    return transpose(v) * m_H * v + 2 * dot(m_c, v) + m_k; 
  }
}
\end{lstlisting}

\subsection{Boundary handling with QEM}

{\small The standard QEM method struggles with boundary edges - those with only one adjacent face. As noted in \cite{Garland1998Simplifying}, a modified QEM was proposed to address this and preserve boundary edges.}

\begin{figure}[!ht]
\centering
   \includegraphics[width=0.65\columnwidth]{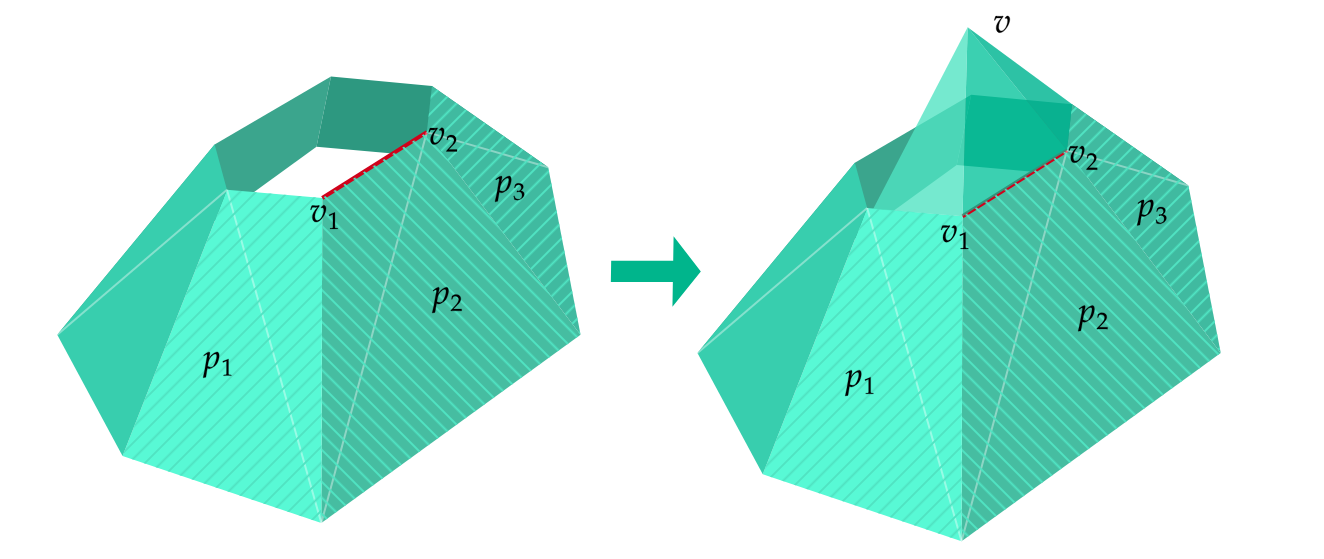}
   \caption{\label{fig:boundary-qem-demo} When a red edge is collapsed, QEM places the new vertex at the intersection of adjacent planes. This can cause the new vertex to be located away from the boundary}
\end{figure}

{\small \autoref{fig:boundary-qem-demo} illustrates the boundary-related limitations of the standard QEM approach. Consider the red boundary edge between $v_{1}$ and $v_{2}$, selected for collapse under two distinct surrounding geometries. The new vertex $v$ is computed as the intersection of the adjacent planes $p_{1} ,\ p_{2}$ and $p_{3}$ because the distance of that point from all these planes is zero - resulting in the minimum possible quadric error. Depending on their configuration, this intersection may lie above or below the original boundary.}

\begin{figure}[!ht]
\centering
   \includegraphics[width=0.65\columnwidth]{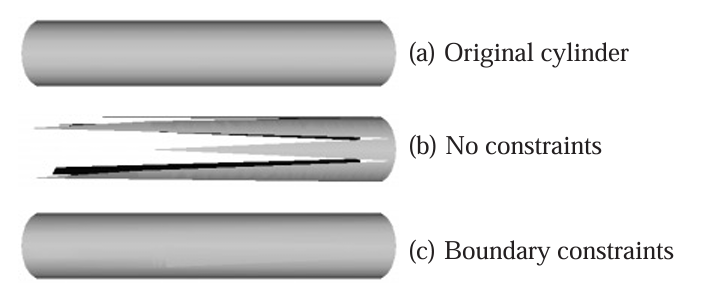}
   \caption{\label{fig:boundary-qem-comparison} Comparison of QEM with and without boundary constraints}
\end{figure}

{\small In conventional QEM, no explicit constraint anchors the new vertex to the boundary. Consequently, collapsing a boundary edge tends to displace the vertex away from the boundary, a deviation that compounds as more boundary edges are collapsed. This progressive drift results in noticeable degradation of mesh quality, as seen in \autoref{fig:boundary-qem-comparison}.}

\begin{figure}[!ht]
\centering
   \includegraphics[width=0.65\columnwidth]{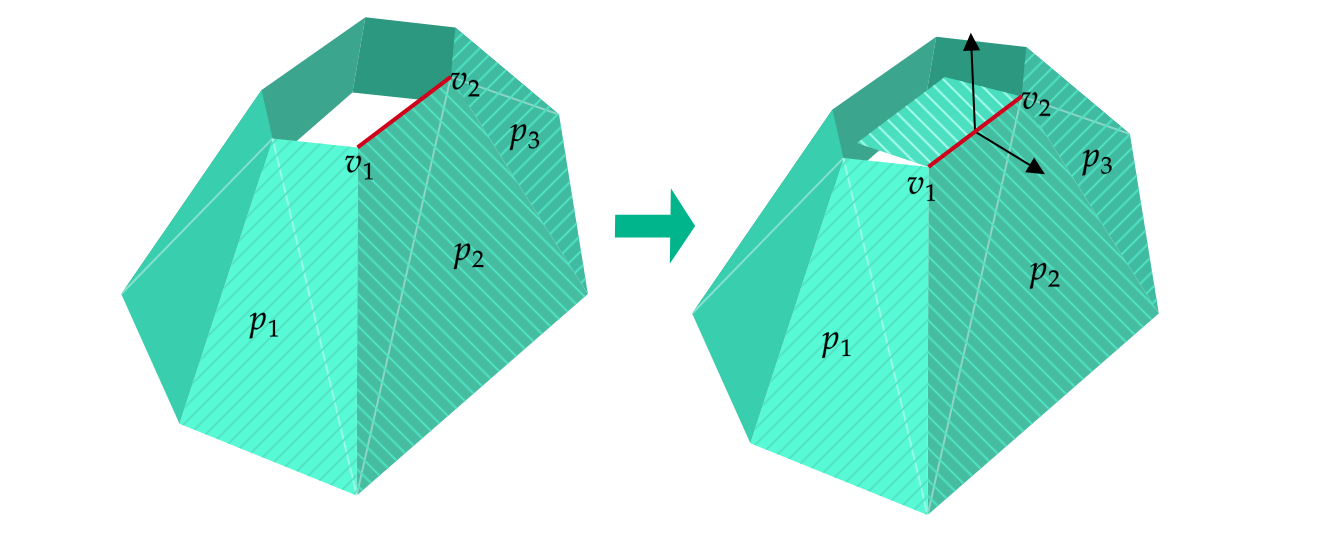}
   \caption{\label{fig:boundary-qem-imaginary-plane} Imaginary plane added for boundary handling}
\end{figure}

{\small To counteract this effect, \cite{Garland1998Simplifying} introduces an imaginary plane $p'$ as shown in \autoref{fig:boundary-qem-imaginary-plane} in addition to the actual planes adjacent to the collapsing edge. $p'$ is defined as the plane perpendicular to the mesh plane containing edge $( v_1 ,v_2)$. A new term, $d^{2}( v,p')$, representing the squared distance between the vertex and $p'$, is incorporated into the error metric. As $v$ moves away from the boundary, this term increases, exerting a corrective pull toward the boundary. To strengthen this constraint, the quadric for $p'$ is scaled by a large constant before being added to the quadrics of the edge endpoints.}

\begin{figure}[!ht]
\centering
   \includegraphics[width=0.55\columnwidth]{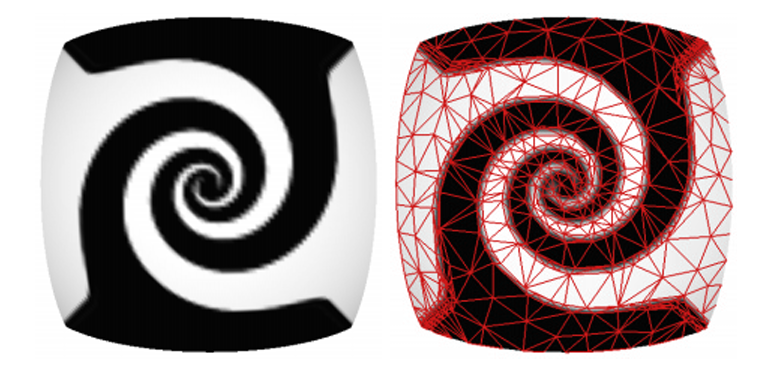}
   \caption{\label{fig:boundary-qem-vertex-colors} A mesh with vertex colors before and after simpification by enhanced QEM}
\end{figure}

{\small The method is further extended to treat edges separating faces with different attribute values (e.g., material indices) as boundaries. This ensures that such attribute boundaries are preserved during simplification, concentrating edges and faces along these divisions for improved alignment. An example of this extension is shown in \autoref{fig:boundary-qem-vertex-colors}.}

\begin{lstlisting}
class BoundaryQEM : QEM
{
  void BoundaryQEM(const IMesh* mesh, const IEdge* collapse_edge) : QEM(mesh, collapse_edge)
  {
    static const double BOUNDARY_QUADRIC_SCALE = 10.0;  // scale factor for quadric of imaginary boundary planes

    auto vertices = collapse_edge->GetVertices();
    auto connected_faces = GetUnion(
      mesh->GetConnectedFaces(vertices[0]),
      mesh->GetConnectedFaces(vertices[1])
    );

    for (auto face : connected_faces) 
    {
      auto face_normal = util::ComputeNormal(face);
      
      for(const auto face_edge : face.GetEdges())
      {
          if(!util::IsBoundaryEdge(face_edge)) continue;
       
          auto edge_verts = face_edge->GetVertices();
          vec3 edge_pos[2] = { edge_verts[0]->GetPosition(), edge_verts[1]->GetPosition() };
          vec3 perp_to_face = normalize(cross(face_normal, edge_pos[1] - edge_pos[0]));

          double d = -dot(perp_to_face, edge_pos[0]);
          m_H += BOUNDARY_QUADRIC_SCALE * outerProduct(perp_to_face, perp_to_face);
          m_c += BOUNDARY_QUADRIC_SCALE * d * perp_to_face;
          k   += BOUNDARY_QUADRIC_SCALE * d * d;
      }
    }
  }
}
\end{lstlisting}

\subsection{Volume preservation constraint}

{\small This constraint, introduced in \cite{Lindstrom1998Turk} helps preserve the mesh volume. If the new vertex replacing the collapsed edge isn’t chosen carefully, it can distort the model. For instance, using the edge midpoint as the new vertex might increase the volume in concave areas or decrease it in convex ones. The goal of this constraint is to preserve volume locally at each collapse, thereby minimizing the overall volume change across the whole model.}

{\small Neither boundary nor volume preservation guarantee geometric integrity; boundaries may deform, and surfaces can lose detail. However, these constraints serve as useful heuristics. Preserving simple, quantifiable properties like area and volume helps reduce extreme distortions, even if local features like sharp edges or curves are lost. While these constraints don't capture fine geometric details, they provide an efficient way to maintain overall structure, balancing accuracy and performance without the complexity of exact boundary or volume preservation.}

\begin{figure}[!ht]
\centering
   \includegraphics[width=0.65\columnwidth]{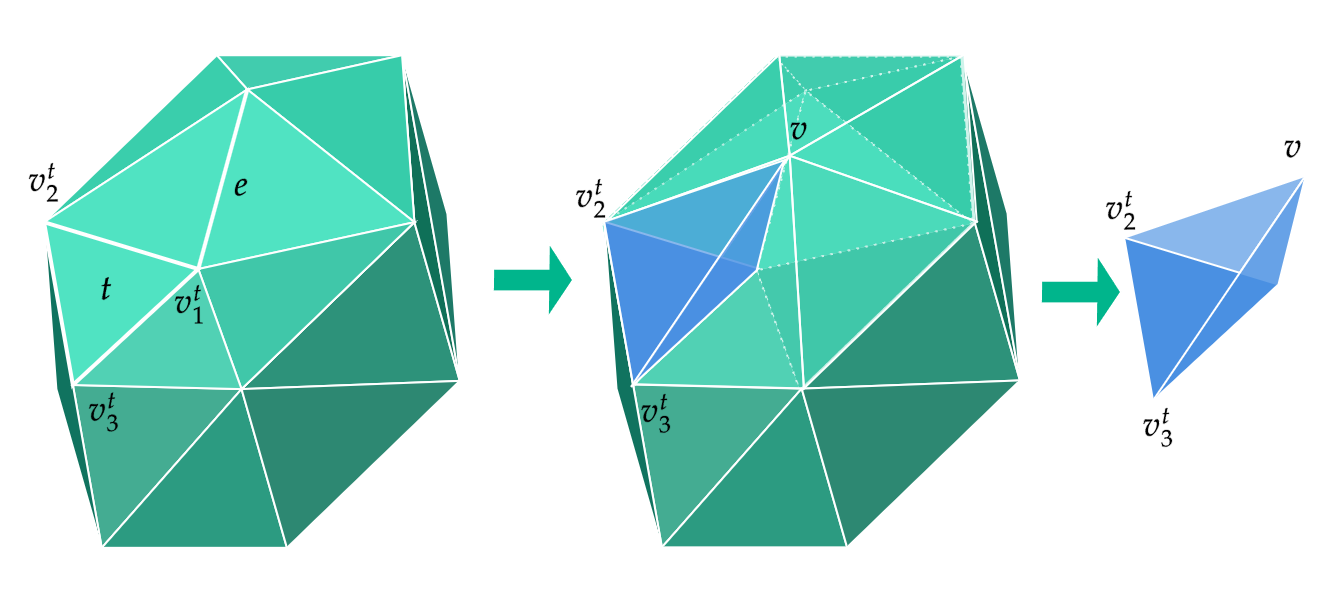}
   \caption{\label{fig:vol-pres-sweep-tet} Volume preservation - sweeping a tetrahedron}
\end{figure}

{\small When an edge $e$ is collapsed, it sweeps out a tetrahedral volume as illustrated in \autoref{fig:vol-pres-sweep-tet}, due to each triangle $t$ being replaced by a new triangle $t'$.}

{\small Let $t=[v_{1},v_{2},v_{3}]$,\ $t'=[v,v_{2},v_{3}]$, and the volume swept by $t$ as$\ v_{1}$ moves linearly to $v$ be $V( v,v_{1} ,v_{2} ,v_{3})$. $V$ is positive if $v$ is above the plane of $t$ and negative otherwise.}

{\small Thus, to preserve the local volume at the site of an edge collapse, the sum of volumes of tetrahedra swept with all triangles $T=\{t_{1} ,\ t_{2} ,\ ...,\ t_{n}\}$ connected to edge $e$ are considered. The change in volume is given by (the superscript $t$ indicates the vertices belonging to triangle $t$):}

{\small \begin{equation*}
\begin{aligned}
{\Eulerconst }\ & {=\sum _{t\in T} V\left( v,\ v_1^t ,\ v_2^t ,\ v_3^t\right)}\\
 & {=\sum _{t\in T}\frac{1}{6}\begin{vmatrix}
v_{x} & v_{1x}^{t} & v_{2x}^{t} & v_{3x}^{t}\\
v_{y} & v_{1y}^{t} & v_{2y}^{t} & v_{3y}^{t}\\
v_{z} & v_{1z}^{t} & v_{2z}^{t} & v_{3z}^{t}\\
1 & 1 & 1 & 1
\end{vmatrix} \ }
\end{aligned}
\end{equation*}}

{\small Solving for $\Eulerconst =0$ and expanding the determinant along the fourth row, we get,}
{\small \begin{equation*}
{0=\frac{1}{6}\sum _{t\in T}\begin{Bmatrix}
\textcolor[rgb]{0.25,0.46,0.02}{-1\cdot }\textcolor[rgb]{0.25,0.46,0.02}{\begin{vmatrix}
v_{1x}^{t} & v_{2x}^{t} & \ v_{3x}^{t}\\
v_{1y}^{t} & v_{2y}^{t} & v_{3y}^{t}\\
v_{1z}^{t} & v_{2z}^{t} & v_{3z}^{t}
\end{vmatrix}}\textcolor[rgb]{0.56,0.07,1}{+1\cdot }\textcolor[rgb]{0.56,0.07,1}{\begin{vmatrix}
v_{x} & v_{2x}^{t} & \ v_{3x}^{t}\\
v_{y} & v_{2y}^{t} & v_{3y}^{t}\\
v_{z} & v_{2z}^{t} & v_{3z}^{t}
\end{vmatrix}}\textcolor[rgb]{0.29,0.56,0.89}{-1\cdot }\textcolor[rgb]{0.29,0.56,0.89}{\begin{vmatrix}
v_{x} & v_{1x}^{t} & \ v_{3x}^{t}\\
v_{y} & v_{1y}^{t} & v_{3y}^{t}\\
v_{z} & v_{1z}^{t} & v_{3z}^{t}
\end{vmatrix}}\textcolor[rgb]{0.74,0.06,0.88}{+1\cdot }\textcolor[rgb]{0.74,0.06,0.88}{\begin{vmatrix}
v_{x} & v_{1x}^{t} & v_{2x}^{t}\\
v_{y} & v_{1y}^{t} & v_{2y}^{t}\\
v_{z} & v_{1z}^{t} & v_{2z}^{t}
\end{vmatrix}}
\end{Bmatrix}}
\end{equation*}}
{\small Representing the determinants that include $v$ as scalar triple products, we get,}
{\small \begin{equation} \label{eq:volume-pres}
\begin{aligned}
{0}\ & {=\ \sum _{t\in T}
\begin{Bmatrix}
\textcolor[rgb]{0.56,0.07,1}{v\cdot \left(v_2^t\times v_3^t\right)} -
\textcolor[rgb]{0.29,0.56,0.89}{v\cdot\left(v_1^t\times v_3^t\right)} +
\textcolor[rgb]{0.74,0.06,0.88}
{
v\cdot\left(v_1^t\times v_2^t\right)
} - 
\textcolor[rgb]{0.25,0.46,0.02}
{
\begin{vmatrix}
v_1^t & v_2^t & v_3^t
\end{vmatrix}
}
\end{Bmatrix}}\\
 & {=v\cdotp \sum _{t\in T}\left(\left( v_2^t \times v_3^t\right) +\left( v_3^t \times v_1^t\right) +\left( v_1^t \times v_2^t\right)\right) -\sum _{t\in T}
\begin{vmatrix}
v_1^t & v_2^t & v_3^t
\end{vmatrix}} \\
\end{aligned}
\end{equation}}
{\small Simplifying the term $\left( v_2^t \times v_3^t +v_3^t \times v_1^t +v_1^t \times v_2^t\right)$, we get:}
{\small \begin{equation*}
\begin{aligned}
\left( v_2^t \times v_3^t +v_3^t \times v_1^t +v_1^t \times v_2^t\right) & =\left( v_2^t \times v_3^t\right) +\left( v_3^t \times v_1^t\right) +\left( v_1^t \times v_2^t\right) \ +\mathbf{\left( v_1^t \times v_1^t\right)} \\
 & =\left( v_2^t \times v_3^t\right) -\left( v_1^t \times v_3^t\right) -\left( v_2^t \times v_1^t\right) +\left( v_1^t \times v_1^t\right) \\
 & =- v_2^t \times \left( v_1^t -v_3^t\right) +v_1^t \times \left( v_1^t -v_3^t\right) \\
 & =\left( v_1^t -v_2^t\right) \times \left( v_1^t -v_3^t\right) \\
 & =\left( v_2^t -v_1^t\right) \times \left( v_3^t -v_1^t\right) \\
 & =n^{t}
\end{aligned}
\end{equation*}}

{\small where $n^{t}$ is the normal of the plane containing $\triangleserifs \left( v_1^t ,v_2^t ,v_3^t\right)$ with magnitude equal to triangle area.}

{\small Substituting the above simplified term in \autoref{eq:volume-pres}, we get,}
{\small \begin{equation*}
\begin{aligned}
0 \  & {= v\cdotp \sum _{t\in T} n^{t} -\sum _{t\in T}\begin{vmatrix}
v_1^t & v_2^t & v_3^t
\end{vmatrix}} \\
{\Longrightarrow v\cdotp \sum _{t\in T} n^{t}} \  & {=\sum _{t\in T}\begin{vmatrix}
v_1^t & v_2^t & v_3^t
\end{vmatrix}}
\end{aligned}
\end{equation*}}

{\small The above equation is of the form $v\cdotp N=D$ which defines a plane. This implies that the vector $v$ is restricted to lie on a plane. So, any point on that plane will satisfy the equation above, implying that volume preservation alone is not enough to fully determine $v$: we need 2 other constraint equations to do so.}

\begin{lstlisting}
class VolumePres : IConstraint
{
  void VolumePres(const IMesh* mesh, const IEdge* collapse_edge)
  {
    m_H = mat3(0);  m_c = vec3(0);  m_k = 0;

    auto vertices = collapse_edge->GetVertices();
    auto connected_faces = GetUnion(
      mesh->GetConnectedFaces(vertices[0]),
      mesh->GetConnectedFaces(vertices[1])
    );

    for (auto face : connected_faces)
    {
      vec3 face_normal = util::ComputeNormal(face);
    
      auto fv = face->GetVertices();
      vec3 positions[3] = { fv[0]->GetPosition(), fv[1]->GetPosition(), fv[2]->GetPosition() };
      mat3 D(0);
      D[0] = positions[0];    D[1] = positions[1];    D[2] = positions[2];
      
      float det = determinant(D);
    
      m_H[0] += face_normal;
      m_c[0] += det;
    }
  }
  
  double EvaluateCost(const vec3& v) const override 
  { 
    return 0; 
  }
}
\end{lstlisting}

\subsection{Volume optimization constraint}

{\small Volume optimization finds the best new vertex position by minimizing the total unsigned volume change during an edge collapse. It fully determines the position, i.e., no extra constraints needed.}

{\small By contrast, volume preservation only ensures that the total volume added and removed balances out to zero. And, as we know, it forces the vertex to lie on a plane but doesn’t tell us exactly where on that plane to place it, so it leaves some freedom. Moreover, it can lead to local distortions if large volumes are added and subtracted in different areas.}

{\small From the volume preservation constraint formulation, we know that the change of volume induced by an edge collapse is:}

{\small \begin{equation*}
{\Eulerconst \ =\sum _{t\in T} V\left( v,\ v_1^t ,\ v_2^t ,\ v_3^t\right)}
\end{equation*}}

{\small where:}

\begin{itemize}
\item {\small $t=\triangleserifs\left(v_1^t,v_2^t,v_3^t\right)$}
\item {\small $t'=\triangleserifs(v,v_2,v_3)$}
\item {\small $V(v,v_1,v_2,v_3)$ is the volume swept out by $t$ when $v_{1}$ moves in a linear path to $v$}
\end{itemize}

{\small If the vertex $v$ is above the plane of a triangle $t$, the signed volume $V$ of the tetrahedron is positive. If below, it's negative. But for optimization, we care about how much the volume changes, not the direction. So, we use the unsigned volume change. To get unsigned volume, we could use $|V|$ or $V^{2}$. We use $V^{2}$ because it is differentiable everywhere. This matters because optimization algorithms rely on gradients and $|V|$ has a kink at zero where the gradient is undefined.}

{\small So we express $\Eulerconst $ as the sum of squares of volumes instead. So we get,}

{\small 
\begin{equation*}
\begin{aligned}
{\Eulerconst } & {=\sum _{t\in T} V\left( v,\ v_1^t ,\ v_2^t ,\ v_3^t\right)^{2}}\\
 & {=\sum _{t\in T}\left( v\cdotp n_{t} -\begin{vmatrix}
v_1^t & v_2^t & v_3^t \end{vmatrix}\right)^{2}}\\
 & {=\sum _{t\in T}\left(\left( v\cdotp n_{t}\right)^{2}
 +\begin{vmatrix}
v_1^t & v_2^t & v_3^t
\end{vmatrix}^{2} -2 v\cdotp n_{t} \begin{vmatrix}
v_1^t & v_2^t & v_3^t
\end{vmatrix}\right)}\\
 & {=\left( \sum _{t\in T}v^{T}\left( n_{t} n_{t}^{T}\right) v\right) -2\left(\sum _{t\in T}\left(\begin{vmatrix}
v_1^t & v_2^t & v_3^t
\end{vmatrix} n_{t}\right) \cdotp v\right) +\sum _{t\in T}\begin{vmatrix}
v_1^t & v_2^t & v_3^t
\end{vmatrix}^{2} \ }\\
 & {=v^{T}\textcolor[rgb]{0.82,0.01,0.11}{\left({\sum _{t\in T}}\left(n_t{n_t}^T\right)\right)} v+2
 \textcolor[rgb]{0.29,0.56,0.89}{\left(- {\sum _{t\in T}}\left({\begin{vmatrix} v_1^t & v_2^t & v_3^t
\end{vmatrix}} n_t\right)\right)} \cdotp v+
\textcolor[rgb]{0.74,0.06,0.88}{\sum _{t\in T}
\begin{vmatrix}
v_1^t & v_2^t & v_3^t
\end{vmatrix}{^2} \ }}\\
 & {=v^{T}\textcolor[rgb]{0.82,0.01,0.11}{H} v+2\textcolor[rgb]{0.29,0.56,0.89}{c}^{T} v+\textcolor[rgb]{0.74,0.06,0.88}{k}}
\end{aligned}
\end{equation*}}

{\small and we get,}

{\small \begin{equation*}
\ v=-H^{-1} c
\end{equation*}}

{\small This constraint uniquely determines $v$, except in degenerate cases where $\det(H) = 0$. Just like the case of QEM, this happens in locally flat regions of the geometry, since we know that $H$ is defined as:}

{\small \begin{equation*}
H=\sum _{t}\left( {n_t} {n_t}^T\right)
\end{equation*}}

{\small which has the form used in QEM. So, $H$ becomes non-invertible in flat regions, where all $n_t$ are parallel and the sum reduces to scaled rank-1 terms. In such situations, alternative constraints or fallback strategies are required for vertex placement.}

\begin{lstlisting}
class VolumeOpt : IConstraint
{
  void VolumeOpt(const IMesh* mesh, const IEdge* collapse_edge)
  {
    auto vertices = collapse_edge->GetVertices();
    auto connected_faces = GetUnion(
      mesh->GetConnectedFaces(vertices[0]),
      mesh->GetConnectedFaces(vertices[1])
    );
    
    m_H = mat3(0);  m_c = vec3(0);  m_k = 0;

    for (const auto face : connected_faces)
    {
      vec3 face_normal = util::ComputeNormal(face);

      auto fv = face->GetVertices();
      vec3 positions[3] = { fv[0]->GetPosition(), fv[1]->GetPosition(), fv[2]->GetPosition() };
      
      mat3 D(0);
      D[0] = positions[0];    D[1] = positions[1];    D[2] = positions[2];
      
      float det = determinant(D);
    
      m_H += outerProduct(face_normal, face_normal);
      m_c += det * face_normal;
      m_k += det * det;
    }

    m_c = -1.0 * m_c;
  }

  double EvaluateCost(const vec3& v) const override 
  { 
    return transpose(v) * m_H * v + 2.0 * dot(m_c, v) + m_k; 
  }
}
\end{lstlisting}

\subsection{Boundary preservation constraint}

{\small This constraint is discussed in \cite{Lindstrom1998Turk}. It helps compute optimal vertex placement and edge collapse error by preserving the area of boundaries.}

\begin{figure}[!ht]
\centering
   \includegraphics[width=0.65\columnwidth]{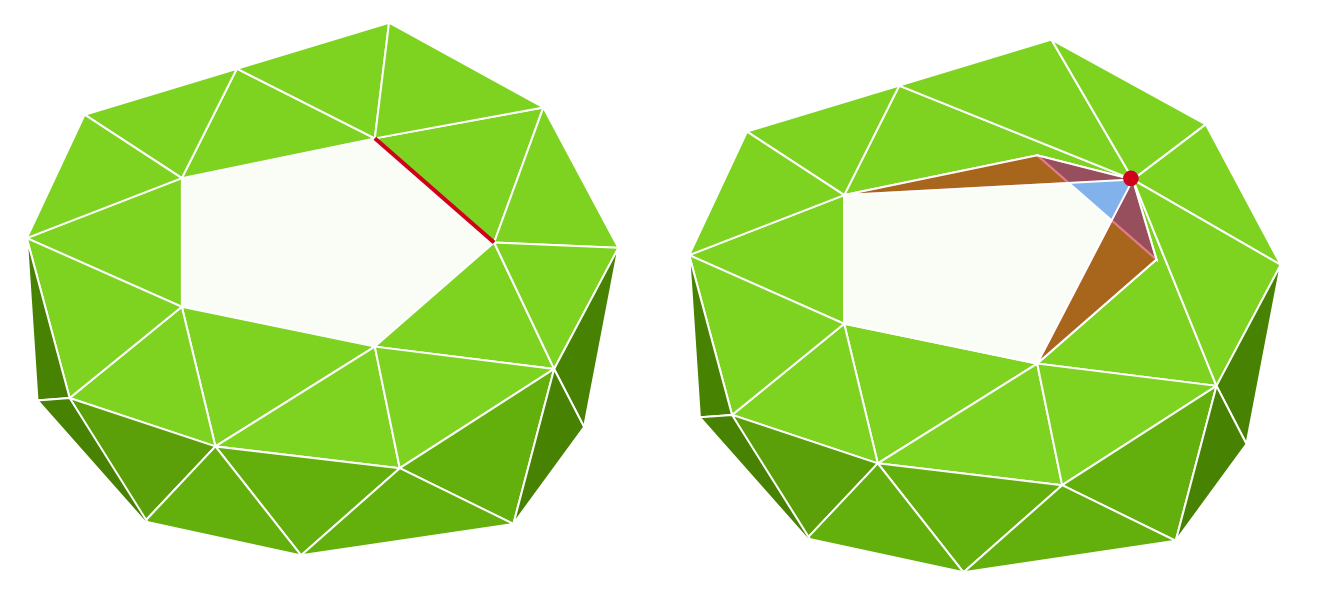}
   \caption{\label{fig:boundary-pres-example} Boundary preservation example (areas not to scale)}
\end{figure}

{\small In \autoref{fig:boundary-pres-example}, the image on the left shows a boundary edge (in red) on a planar hole, while the image on the right shows it being replaced by a new vertex (red dot) after edge collapse. As a result, although the total shaded area is preserved (red area loss offset by blue area gain), the boundary’s shape and structure are visibly altered. Thus, the constraint preserves area, not the boundary itself.}

\begin{figure}[!ht]
\centering
   \includegraphics[width=0.65\columnwidth]{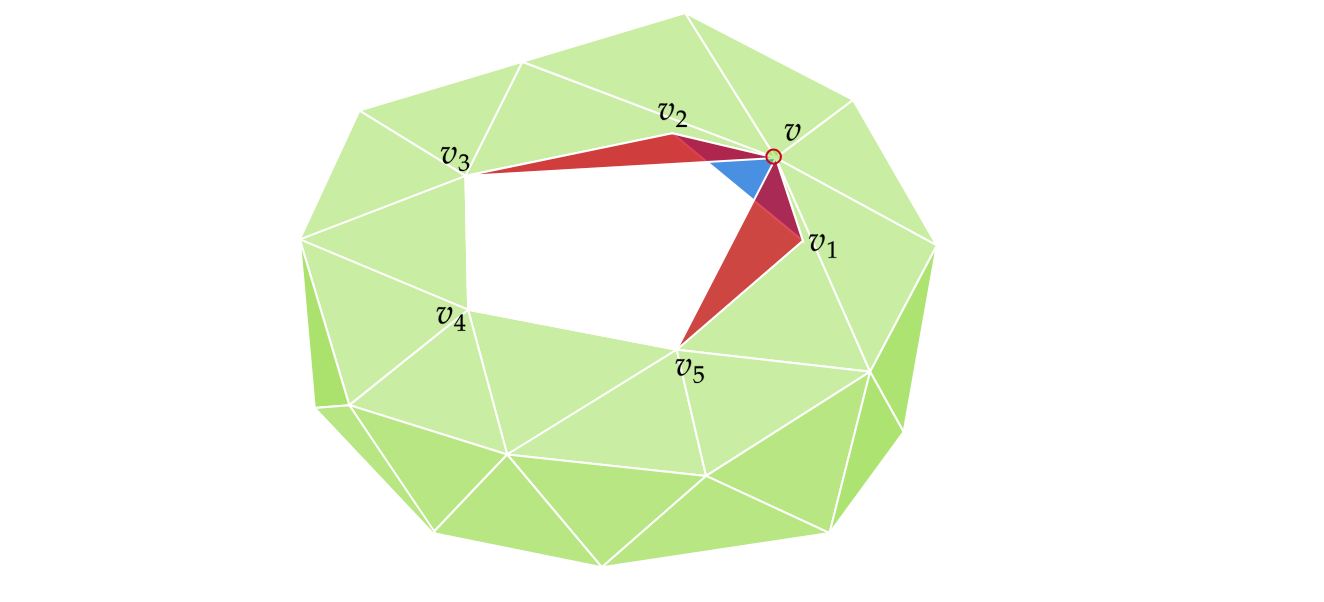}
   \caption{\label{fig:boundary-pres-planar} Boundary preservation - planar boundary case}
\end{figure}

{\small As per \autoref{fig:boundary-pres-planar}, let edge $\displaystyle \left( v_1,v_2\right)$ be collapsed into vertex $\displaystyle v$ and let $\displaystyle \Eulerconst $ be the net area change. Collapsing a boundary edge connects the new vertex $\displaystyle v$ to two other boundary edges, forming three triangles. The sum of their signed areas gives yields $\displaystyle \Eulerconst $.}

{\small \begin{equation*}
\begin{aligned}
\Eulerconst  & =\textcolor[rgb]{0.29,0.56,0.89}{Area}\textcolor[rgb]{0.29,0.56,0.89}{(}\textcolor[rgb]{0.29,0.56,0.89}{v,v}\textcolor[rgb]{0.29,0.56,0.89}{_{1}}\textcolor[rgb]{0.29,0.56,0.89}{,v}\textcolor[rgb]{0.29,0.56,0.89}{_{2}}\textcolor[rgb]{0.29,0.56,0.89}{)} \ +\ \textcolor[rgb]{0.82,0.01,0.11}{Area}\textcolor[rgb]{0.82,0.01,0.11}{(}\textcolor[rgb]{0.82,0.01,0.11}{v,v}\textcolor[rgb]{0.82,0.01,0.11}{_{2}}\textcolor[rgb]{0.82,0.01,0.11}{,v}\textcolor[rgb]{0.82,0.01,0.11}{_{3}}\textcolor[rgb]{0.82,0.01,0.11}{)} \ +\ \textcolor[rgb]{0.82,0.01,0.11}{Area}\textcolor[rgb]{0.82,0.01,0.11}{(}\textcolor[rgb]{0.82,0.01,0.11}{v,v}\textcolor[rgb]{0.82,0.01,0.11}{_{5}}\textcolor[rgb]{0.82,0.01,0.11}{,v}\textcolor[rgb]{0.82,0.01,0.11}{_{1}}\textcolor[rgb]{0.82,0.01,0.11}{)}\\
 & =\frac{1}{2}\left(\begin{aligned}
\left( v\times v_{1}\right) +\left( v_{1} \times v_{2}\right) +\left( v_{2} \times v\right) & +\\
\left( v\times v_{2}\right) +\left( v_{2} \times v_{3}\right) +\left( v_{3} \times v\right) & +\\
\left( v\times v_{5}\right) +\left( v_{5} \times v_{1}\right) +\left( v_{1} \times v\right) & \end{aligned}\right)
\end{aligned}
\end{equation*}}

{\small While planar boundaries help build intuition for area change during edge collapse, real boundaries are often non-planar. However, the same formulation works for non-planar boundaries, as detailed in the appendix (\autoref{sec:appendix}).}

\begin{figure}[!ht]
\centering
   \includegraphics[width=0.65\columnwidth]{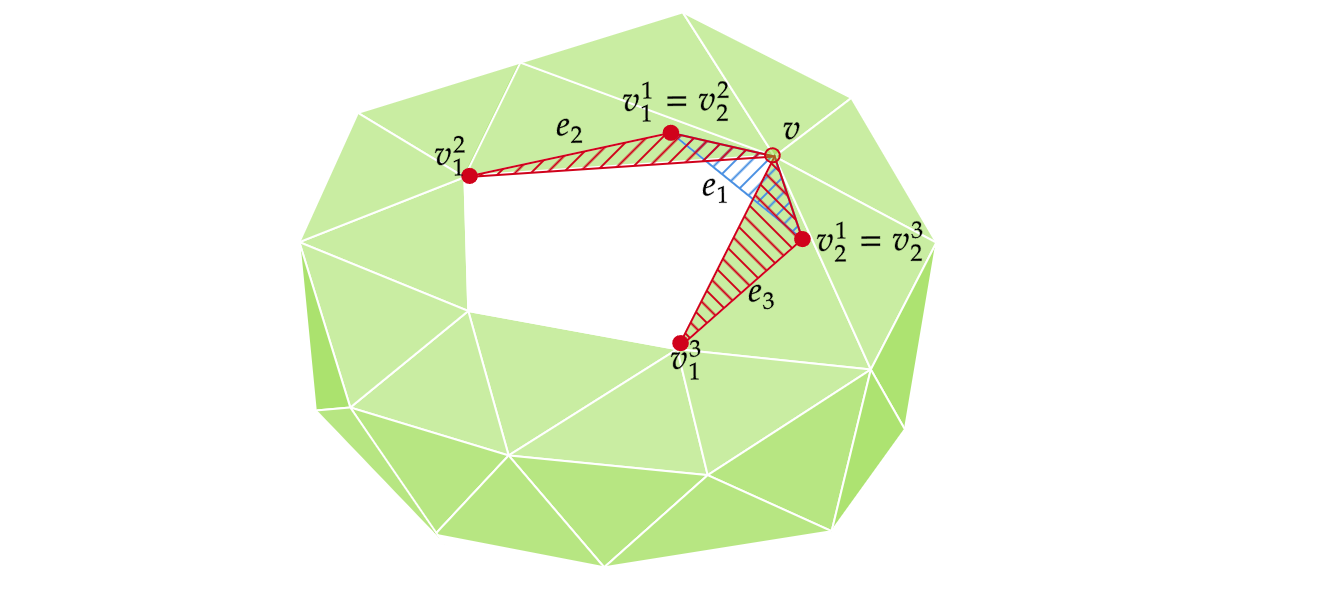}
   \caption{\label{fig:boundary-pres-constraint-derivation} Boundary preservation - constraint derivation}
\end{figure}

{\small Let $\displaystyle \Eulerconst$ be the squared change in area from the edge collapse, and $\displaystyle E=\{e_1,\dotsc ,e_n\}$ the set of boundary edges connected to vertex $\displaystyle v$ with each edge $\displaystyle e:\left( v_1^e,v_2^e\right) \in E$.}

{\small Collapsing a boundary edge connects two others, forming three triangles ($n=3$) as shown in \autoref{fig:boundary-pres-constraint-derivation}. The squared norm is used for computational simplicity. The error is then defined as the squared magnitude of the total area change induced by vertex $\displaystyle v$ as:}

{\small \begin{equation*}
\begin{aligned}
\Eulerconst & =\begin{Vmatrix}
{\textstyle \sum _{e\in E}} \ Area\left( v,\ v_1^e ,\ v_2^e\right)
\end{Vmatrix}^{2} \\
& =\begin{Vmatrix}
{\textstyle \sum _{e\in E}}\frac{1}{2}\left(\left( v\times v_1^e\right) +\left( v_1^e \times v_2^e\right) +\left( v_2^e \times v\right)\right)
\end{Vmatrix}^{2} \\
 & =\begin{Vmatrix}
{\textstyle \sum _{e\in E}}\frac{1}{2}\left( v\times \left( v_1^e -v_2^e\right) +\left( v_1^e \times v_2^e\right)\right)
\end{Vmatrix}^{2} \\
 & =\frac{1}{4}\begin{Vmatrix}
v\times {\textstyle \sum _{e\in E}\left( v_1^e -v_2^e\right) +\sum _{e\in E}\left( v_1^e \times v_2^e\right)}
\end{Vmatrix}^{2} \\
\end{aligned}
\end{equation*}}

{\small For simplicity, denote $\displaystyle {\textstyle \sum _{e\in E}\left( v_1^e - v_2^e\right)}$ as $\displaystyle E_1$ and $\displaystyle {\textstyle \sum _{e\in E}\left( v_1^e \times v_2^e\right)}$ as $\displaystyle E_{2}$. This gives,}

{\small \begin{equation*}
\Eulerconst = \frac{1}{4}\begin{Vmatrix}
v\times E_{1} +E_{2}
\end{Vmatrix}^{2} = \frac{1}{4}\begin{Vmatrix}
E_{2} -E_{1} \times v
\end{Vmatrix}^{2}
\end{equation*}}

{\small To simplify the cross-product term, $\displaystyle E_{1} \times v$ can be written as $\displaystyle \mathcal{S} v$, where $\displaystyle \mathcal{S}$ is the skew-symmetric matrix for the cross product with $\displaystyle E_{1}$:}

{\small \begin{equation*}
\Eulerconst =\ \frac{1}{4}\begin{Vmatrix}
E_{2} -\mathcal{S} v
\end{Vmatrix}^{2}
\end{equation*}}

{\small Simplifying the squared norm gives:}

{\small \begin{equation*}
\begin{aligned}
\Eulerconst  & =\frac{1}{4}( E_{2} -\mathcal{S} v)^{T}( E_{2} -\mathcal{S} v)\\
 & =\frac{1}{4}\left( E_{2}^{T} -v^{T}\mathcal{S}^{T}\right)( E_{2} -\mathcal{S} v)\\
 & =\frac{1}{4}\left( v^T\textcolor[rgb]{0.82,0.01,0.11}{\mathcal{S}^T\mathcal{S}} v+2\left(\textcolor[rgb]{0.29,0.56,0.89}{- {E_2}^T \mathcal{S}}\right)^{T} v +\textcolor[rgb]{0.74,0.06,0.88}{E_2^T E_2}\right)\\
 & =\frac{1}{4}\left( v^{T}\textcolor[rgb]{0.82,0.01,0.11}{H} v+2\textcolor[rgb]{0.29,0.56,0.89}{c}^{T} v+\textcolor[rgb]{0.74,0.06,0.88}{k}\right)
\end{aligned}
\end{equation*}}
{\small which has the same form as in QEM, suggesting that the solution should be the same: $\displaystyle v=-H^{-1} c$.}

{\small However, for this constraint, $\displaystyle H$ is non-invertible as $\displaystyle \mathcal{S}$ is non-invertible, being a skew-symmetric matrix. Therefore, $\displaystyle v$ cannot be fully determined. Below is a demonstration of the constraints that can actually be extracted from setting the gradient to zero.}

{\small \begin{equation*}
\begin{aligned}
\nabla \Eulerconst & =\frac{1}{4}( 2Hv+2c) & \\
\Longrightarrow 0 & =\frac{1}{4}\left( 2\mathcal{S}^{T}\mathcal{S} v+2\mathcal{S} E_{2}\right) & \dotsc \ \mathcal{S}^{\mathtt{T}} =-\mathcal{S}\text{ for skew-symmetric matrices}\\
\Longrightarrow 0 & =\mathcal{S} E_{2} -\mathcal{S}^{2} v & \\
 & =\mathcal{S}( E_{2} - \mathcal{S}v ) & \dotsc \ \mathcal{S}\text{ cannot be cancelled with 0 as it is non-invertible}\\
 & =E_{1} \times ( \ E_{2} -E_{1} \times v) & \\
 & =E_{1} \times E_{2} -( E_{1} \times ( E_{1} \times v)) \  & \\
 & =E_{1} \times E_{2} -E_{1}( E_{1} \cdotp v) +v( E_{1} \cdotp E_{1}) & ...\text{ using vector triple product}\\
 & =v\left( \| {\textstyle E_{1}} \| ^{2}\right) -{\textstyle E_{1}}({\textstyle E_{1}} \cdotp v) +\ K & ...\text{ denoting } E_{1} \times E_{2}\text{ as }K
\end{aligned}
\end{equation*}}

{\small This gives us,}

{\small \begin{equation}\label{eq:boundary-pres-rank2}
0=v\left( \| {\textstyle E_{1}} \| ^{2}\right) -{\textstyle E_{1}}({\textstyle E_{1}} \cdotp v) +{\textstyle K }
\end{equation}}

{\small This is a 3D vector equation in $\displaystyle v$, which can be split into 3 scalar equations to solve for its components. We can choose any basis for this, but for convenience, we choose: $\displaystyle [{\textstyle K ,\ E_{1} ,\ } E_{1} \times {\textstyle K }]$. }\newline

\noindent{\small \textbf{Projecting \autoref{eq:boundary-pres-rank2} onto }$K$\textbf{ gives us:}}
{\small \begin{equation*}
{\textstyle \begin{aligned}
0 & {\textstyle =K \cdotp [} v({\textstyle E_{1}} \cdotp {\textstyle E_{1}}) -{\textstyle E_{1}}({\textstyle E_{1}} \cdotp v) +\ K ] & \\
 & {\textstyle =K \cdotp } v\left( \| {\textstyle E_{1}} \| ^{2}\right) -{\textstyle K \cdotp E_{1}}({\textstyle E_{1}} \cdotp v) +\ {\textstyle K \cdotp } K  & \\
 & =\left[K \cdotp v\left( \| {\textstyle E_{1}} \| ^{2}\right)\right] - \cancel{\mathbf{{\textstyle [ K\cdotp E_{1}}({\textstyle E_{1}} \cdotp v)]}} +\ {\textstyle K \cdotp } K  & ...\ K= {\textstyle ( E_{1}} \times {\textstyle E_{2}) \ \perp E_{1} \ so\ K.E_{1} =0}\\
 & =({\textstyle E_{1}} \times E_{2}) \cdotp \left( v\left( \| {\textstyle E_{1}} \| ^{2}\right)\right) +\ \| {\textstyle E_{1}} \times E_{2} \| ^{2} \ \ \ \ \  & \\
 & =\| {\textstyle E_{1}} \| ^{2}({\textstyle E_{1}} \times E_{2}) \cdotp v+\ \| {\textstyle E_{1}} \times E_{2} \| ^{2} & 
\end{aligned}}
\end{equation*}}

\noindent{\small \textbf{Projecting \autoref{eq:boundary-pres-rank2} onto }$\displaystyle E_{1}$\textbf{ gives us:}}
{\small \begin{equation*}
\begin{aligned}
0 & {\textstyle =E_{1} \cdotp [} v({\textstyle E_{1}} \cdotp {\textstyle E_{1}}) -{\textstyle E_{1}}({\textstyle E_{1}} \cdotp v) +\ K ] & \\
 & =\cancel{\mathbf{{\textstyle E_{1} \cdotp } v({\textstyle E_{1}} \cdotp {\textstyle E_{1}})}} -\cancel{\mathbf{{\textstyle E_{1} \cdotp E_{1}}({\textstyle E_{1}} \cdotp v)}} +\ {\textstyle E_{1} \cdotp ( E_{1} \times E_{2})} & \\
 & {\textstyle =E_{1} \cdotp ( E_{1} \times E_{2})} & \\
 & =0 & {\textstyle ...\ E_{1}} \perp {\textstyle ( E_{1}} \times {\textstyle E_{2})} \ \ 
\end{aligned}
\end{equation*}}
{\small This simplifies to $\displaystyle 0=0$, which is a degenerate result. This indicates that the component of $\displaystyle v$ parallel to $\displaystyle {\textstyle E_{1}}$is not determined by this minimization problem, as it does not affect the value of the error function.}\newline

\noindent{\small \textbf{Projecting \autoref{eq:boundary-pres-rank2} onto }$\displaystyle E_{1} \times {\textstyle K }$\textbf{ gives us:}}
{\small \begin{equation*}
\begin{aligned}
0 & {\textstyle =( E_{1} \times K ) \cdotp [} v({\textstyle E_{1}} \cdotp {\textstyle E_{1}}) -{\textstyle E_{1}}({\textstyle E_{1}} \cdotp v) +\ K ] \  & \\
 & {\textstyle =( E_{1} \times K ) \cdotp } v\left( \| {\textstyle E_{1}} \| ^{2}\right) -{\textstyle ( E_{1} \times K ) \cdotp E_{1}}({\textstyle E_{1}} \cdotp v) +\mathbf{\cancel{\ {\textstyle ( E_{1} \times K ) \cdotp } K \ } \ } & ...\ K \perp {\textstyle ( E_{1} \times K )}\\
 & {\textstyle =( E_{1} \times K ) \cdotp } v\left( \| {\textstyle E_{1}} \| ^{2}\right) -\mathbf{\cancel{{\textstyle ( E_{1} \times K ) \cdotp E_{1}}({\textstyle E_{1}} \cdotp v) \ \ }} & ...\ E_{1} \perp {\textstyle ( E_{1} \times K )}\\
 & {\textstyle =( E_{1} \times ( E_{1} \times E_{2})) \cdotp } v\left( \| {\textstyle E_{1}} \| ^{2}\right) & \\
 & {\textstyle =( E_{1} \times ( E_{1} \times E_{2})) \cdotp } v & 
\end{aligned}
\end{equation*}}
{\small Thus, the solution space of this optimization lies in the intersection of two planes defined by:}
{\small \begin{equation*}
\begin{aligned}
\| {\textstyle E_{1}} \| ^{2}({\textstyle E_{1}} \times {\textstyle E_{2}}) \cdotp v+\ \| {\textstyle E_{1}} \times {\textstyle E_{2}} \| ^{2} =0\\
{\textstyle ( E_{1} \times ( E_{1} \times E_{2}))} \cdotp v=0
\end{aligned}
\end{equation*}}
{\small Even with this approach of minimizing $\displaystyle \Eulerconst $, $\displaystyle v$ remains undetermined. Thus, additional constraints are needed alongside the two equations to solve for $\displaystyle v$.}

\begin{lstlisting}
class BoundaryPres : IConstraint
{
  void BoundaryPres(const IMesh* mesh, const IEdge* collapse_edge)
  {
    if(!util::IsBoundaryEdge(collapse_edge)) return;
  
    m_H = mat3(0);  m_c = vec3(0);  m_k = 0;

    auto vertices = collapse_edge->GetVertices();
    auto connected_edges = GetUnion(
      mesh->GetConnectedEdges(vertices[0]),
      mesh->GetConnectedEdges(vertices[1])
    );

    vec3 E1(0); vec3 E2(0);
    for (auto edge : connected_edges)
    {
      if(!util::IsBoundaryEdge(edge)) continue;

      const auto verts = edge->GetVertices();
      vec3 positions[2] = { verts[0]->GetPosition(), verts[1]->GetPosition() };
    
      E1 += positions[0] - positions[1];
      E2 += cross(positions[0], positions[1]);
    }
    
    mat3 skew_sym_mat = util::MakeSkewSymMat(E1);

    H += transpose(skew_sym_mat) * skew_sym_mat;
    c += skew_sym_mat * E2;
    k += dot(E2, E2);
  }

  double EvaluateCost(const vec3& v) const override 
  { 
    return 0.25 * (transpose(v) * m_H * v) + 0.5 * dot(m_c, v) + 0.25 * m_k; 
  }
}
\end{lstlisting}

\subsection{Boundary Optimization Constraint}

{\small The boundary optimization constraint introduced in \cite{Lindstrom1998Turk} minimizes boundary triangle area like boundary preservation, but focuses on unsigned area. The error is expressed as a sum of squared signed areas, giving:}

{\small \begin{equation}\label{eq:boundary-opt}
\Eulerconst \ ={\textstyle \sum _{e\in E} A\left( v,\ v_1^e ,\ v_2^e\right)}^{2}
\end{equation}}
{\small where,}
\begin{itemize}
    \item {\small $v\ \in \mathbb{R}^3\ $}
    \item {\small $v_1^e ,\ v_2^e =\text{endpoints of each boundary edge } e\in E$}
    \item {\small $A\left( v,\ v_1^e ,\ v_2^e\right) =\text{area formed by } \triangleserifs \left( v,\ v_1^e ,\ v_2^e\right)$}
\end{itemize}

\noindent{\small Expanding \autoref{eq:boundary-opt} gives:}
{\small \begin{equation*}
{\textstyle \Eulerconst \ =\frac{1}{4}\sum _{e\in E}\begin{Vmatrix}
v\times \left( v_1^e -v_2^e\right) +\left( v_1^e \times v_2^e\right)
\end{Vmatrix}^{2}}
\end{equation*}}
{\small Substitute $\displaystyle e_{1} =\left( v_1^e -v_2^e\right)$ and $\displaystyle e_{2} =\left( v_1^e \times v_2^e\right)$, gives:}
{\small \begin{equation*}
\begin{aligned}
\Eulerconst  & {\textstyle =\frac{1}{4}\sum _{e\in E}\begin{Vmatrix}
v\times e_{1} +e_{2}
\end{Vmatrix}^{2} \ } \\
 & {\textstyle =\frac{1}{4}\sum _{e\in E}\left(\begin{Vmatrix}
v\times e_{1}
\end{Vmatrix}^{2} +\begin{Vmatrix}
e_{2}
\end{Vmatrix}^{2} +2( v\times e_{1}) \cdot e_{2}\right)} \\
 & {\textstyle =\frac{1}{4}\left(\sum _{e\in E}\left(\begin{Vmatrix}
v\times e_{1}
\end{Vmatrix}^{2} +2( v\times e_{1})^{T} e_{2}\right) +\sum _{e\in E}\begin{Vmatrix}
e_{2}
\end{Vmatrix}^{2}\right)}
\end{aligned}
\end{equation*}}
{\small To simplify the cross-product term, $\displaystyle v\times e_{1}$ can be written as $\displaystyle \mathcal{S} v$, where $\displaystyle \mathcal{S}$ is the skew-symmetric matrix for the cross product with $\displaystyle e_{1}$. Also, the scalar triple product identity can be used to rearrange $\displaystyle 2( v\times e_{1})^{T} e_{2} =2( e_{1} \times e_{2})^{T} v$.}
{\small \begin{equation*}
\begin{aligned}
{\textstyle \Eulerconst}\ & {\textstyle =\frac{1}{4}\left(\sum _{e\in E}\left(\begin{Vmatrix}
\mathcal{S} v
\end{Vmatrix}^{2} +2( e_{1} \times e_{2})^{T} v\right) +\sum _{e\in E}\begin{Vmatrix}
e_{2}
\end{Vmatrix}^{2}\right)}\\
 & {\textstyle =\frac{1}{4}\left(\sum _{e\in E}\left( v^{T}\left(\mathcal{S}^{T}\mathcal{S}\right) v\right) +2\left(\sum _{e\in E}( e_{1} \times e_{2})^{T}\right) v+\sum _{e\in E}\begin{Vmatrix}
e_{2}
\end{Vmatrix}^{2}\right)}\\
 & {\textstyle =\frac{1}{4}\left( v^{T}\textcolor[rgb]{0.82,0.01,0.11}{\left({\sum _{e\in E}}{\mathcal{S}}{^{T}}{\mathcal{S}}\right)} v+2
 \textcolor[rgb]{0.29,0.56,0.89}{\left({\sum _{e\in E}}(e_1\times e_2)\right)}^{T} v+
 \textcolor[rgb]{0.74,0.06,0.88}{\sum _{e\in E}{\begin{Vmatrix} e_{2} \end{Vmatrix}{^2}}}\right)}\\
 & {\textstyle =\frac{1}{4}\left( v^{T}\textcolor[rgb]{0.82,0.01,0.11}{H} v+2\textcolor[rgb]{0.29,0.56,0.89}{c}^{T} v+\textcolor[rgb]{0.74,0.06,0.88}{k}\right)}
\end{aligned}
\end{equation*}}
{\small which is of the same form as earlier constraints, so we obtain $\displaystyle v$ as $\displaystyle v=-H^{-1} c$.}

{\small As in earlier constraints, if $\displaystyle \det( H) =0$, the constraint becomes degenerate and we need to use other constraints or use fallback strategies for vertex placement.}

\begin{lstlisting}
class BoundaryOpt : IConstraint
{
  void BoundaryOpt(const Mesh* mesh, const IEdge* collapse_edge)
  {
    if(!util::IsBoundaryEdge(collapse_edge)) return;

    auto vertices = collapse_edge->GetConnectedVertices();
    auto connected_edges = GetUnion(
      mesh->GetConnectedEdges(vertices[0]),
      mesh->GetConnectedEdges(vertices[1])
    );
  
    m_H = mat3(0); m_c = vec3(0); m_k = 0;
  
    for (auto edge : connected_edges)
    {
      if(!util::IsBoundaryEdge(edge)) continue;

      const auto verts = edge->GetVertices();
      vec3 positions[2] = { verts[0]->GetPosition(), verts[1]->GetPosition()};
    
      vec3 e1 = positions[0] - positions[1];
      vec3 e2 = cross(positions[0], positions[1]);
    
      mat3 skew_sym_mat = util::MakeSkewSymMat(E1);
    
      m_H += transpose(skew_sym_mat) * skew_sym_mat;
      m_c += cross(e1, e1);
      m_k += dot(e2, e2);
    }
  }

  double EvaluateCost(const vec3& v) const override 
  { 
    return 0.25 * (transpose(v) * m_H * v) + 0.5 * dot(m_c, v) + 0.25 * m_k; 
  }
}
\end{lstlisting}

\subsection{Triangle Shape Optimization Constraint}

{\small Triangle shape optimization tries to improve triangle quality. Skinny or stretched triangles can cause shading issues, while more even, equilateral triangles make the mesh look and work better.}

\begin{figure}[!ht]
\centering
   \includegraphics[width=0.65\columnwidth]{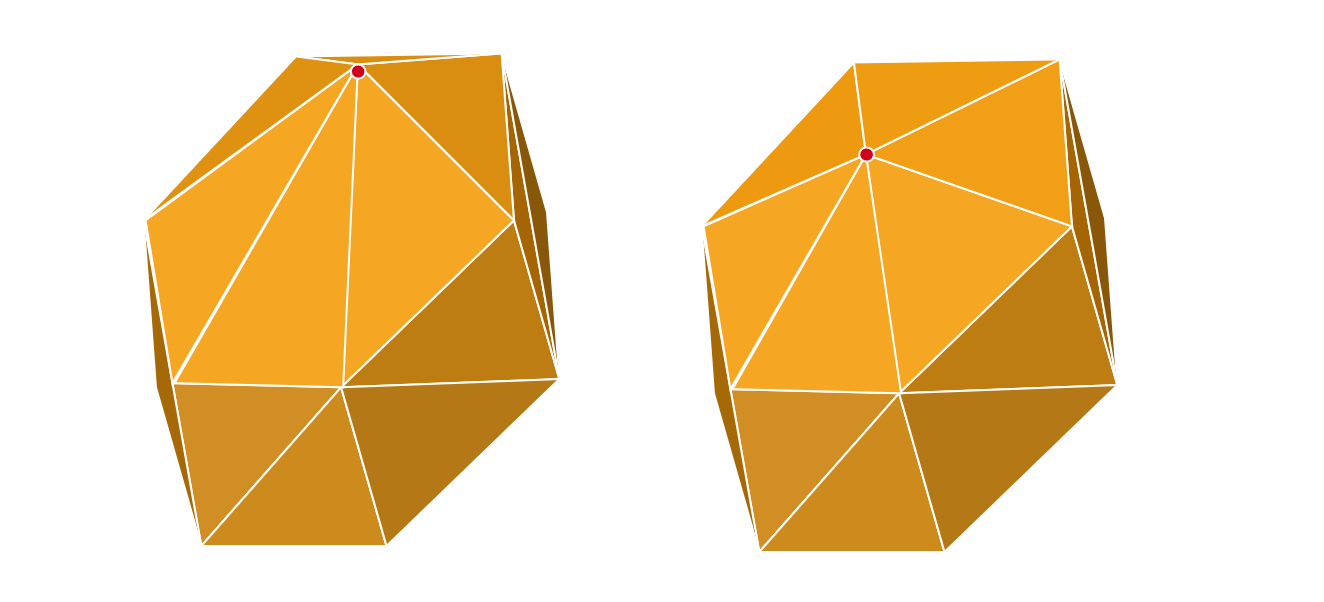}
   \caption{\label{fig:triangle-opt} Edge collapse yielding low-quality (left) vs. high-quality (right) triangle shapes}
\end{figure}

{\small As shown in \autoref{fig:triangle-opt}, the vertex placement on the left side leads to a cleaner triangle structure thanks to its more regular, evenly shaped triangles. The placement on the right side contains long, stretched triangles, which make the mesh look less tidy and visually less appealing.}

{\small Before analyzing this further, an important triangle shape quality metric needs to be introduced: the \textbf{area-to-perimeter ratio}. Regular triangles (like equilateral ones) have a higher area-to-perimeter ratio, which means they're more compact and less stretched.}

{\small In the triangle shape preservation constraint, the goal is to maximize the area-to-perimeter ratio. We make an assumption that the region around the collapsing edge is nearly flat. This means the total area of the nearby triangles doesn’t change much after the edge collapse. So, to improve their area-to-perimeter ratio, we can focus on reducing the perimeter alone, which is determined by the edge lengths.}

{\small That’s why we minimize the sum of the squared lengths of the edges connected to the new vertex. This pulls the vertex into a position where the edges are more evenly distributed and shorter, leading to more balanced, less skinny triangles.}

{\small We formulate this error $\displaystyle \Eulerconst \ $as:}

{\small \begin{equation*}
{\textstyle \Eulerconst \ =\sum _{i=1}^{n} ||v\ -v_{i} ||^{2}}
\end{equation*}}
{\small where $\displaystyle v_{i}$ refers to each neighboring vertex $\displaystyle \{v_{1} ,\dotsc ,v_{n}\}$ connected to$\displaystyle \ v$ in the mesh.}

{\small On expanding this equation for $\displaystyle \Eulerconst $, we get,}
{\small \begin{equation*}
\begin{aligned}
{\textstyle \Eulerconst } & {\textstyle =\sum _{i=1}^{n} ||v\ -v_{i} ||^{2}} & \\
 & {\textstyle =\sum _{i=1}^{n}\left( ||v||^{2} +||v_{i} ||^{2} -2v\cdotp v_{i}\right)} & \\
 & {\textstyle =\sum _{i=1}^{n} ||v||^{2} +\sum _{i=1}^{n} ||v_{i} ||^{2} -2v\cdotp \sum _{i=1}^{n} v_{i}} & \\
 & ={\textstyle \sum _{i=1}^{n} v^{T} v+\sum _{i=1}^{n} ||v_{i} ||^{2} -2v\cdotp \sum _{i=1}^{n} v_{i}} & \\
 & ={\textstyle v^{T}
 \textcolor[rgb]{0.82,0.01,0.11}{\left({\sum _{i=1}^{n}}I\right)} v+2v\cdotp \textcolor[rgb]{0.29,0.56,0.89}{\left(- {\sum _{i=1}^{n}}v_i\right)} +\textcolor[rgb]{0.74,0.06,0.88}{\sum _{i=1}^{n}{||v_i||}{^{2}}}} & \dotsc \ I\text{ is an identity matrix}\\
 & ={\textstyle v^{T}\textcolor[rgb]{0.82,0.01,0.11}{H} v+2\textcolor[rgb]{0.29,0.56,0.89}{c}^{T} v+\textcolor[rgb]{0.74,0.06,0.88}{k}} & 
\end{aligned}
\end{equation*}}

{\small which can be solved in the same way as in other constraints. Here, it leads to this solution:}

{\small \begin{equation*}
v=-H^{-1} c=-\left(\frac{1}{n} I\right){\textstyle \left( -\sum _{i=1}^{n} v_{i}\right) =}\frac{1}{n}{\textstyle \sum _{i=1}^{n} v_{i}}
\end{equation*}}

{\small  Thus, optimizing triangle shape will lead to choosing the centroid of the neighboring vertices.}

{\small Skinny triangles are avoided because they cause shading artifacts. Rasterization interpolates per-vertex data (e.g., normals) using barycentric coordinates, defined as:}

{\small \begin{equation*}
\begin{aligned}
u\ =\ \frac{Area( \triangleserifs ABP)}{Area( \triangleserifs ABC)}, & \ \ \ \ v\ =\ \frac{Area( \triangleserifs APC)}{Area( \triangleserifs ABC)}, & w\ =\ \frac{Area( \triangleserifs BCP)}{Area( \triangleserifs ABC)}
\end{aligned}
\end{equation*}}

{\small In skinny triangles, the denominator $\displaystyle Area( \triangleserifs ABC)$ becomes very small, making the coordinates numerically unstable. Small floating-point errors in the vertex positions or sub-areas can then cause large interpolation errors, producing shading artifacts.}

\begin{lstlisting}
class TriShapeOpt : IConstraint
{
  void TriShapeOpt(const IMesh* mesh, const IEdge* collapse_edge)
  {
    auto vertices = collapse_edge->GetVertices();
    auto connected_verts = GetUnion(
      mesh->GetConnectedVertices(vertices[0]),
      mesh->GetConnectedVertices(vertices[1])
    );

    m_H = mat3(0); m_c = vec3(0);  m_k = 0;

    for (const auto vert : connected_verts)
    {
      m_H += mat3(1); //Identity matrix
      vec3 pos = vert.GetPosition();
      m_c += -pos;
      m_k += dot(pos, pos);
    }
  }

  double EvaluateCost(const vec3& v) const override 
  { 
    return transpose(v) * m_H * v + 2.0 * dot(m_c, v) + m_k; 
  }
}
\end{lstlisting}

\subsection{Fallback strategies}

{\small When the matrix used to compute $v$ is non-invertible, fallback strategies are needed. A simple and common fallback is to place the new vertex at the midpoint of the edge $( v_{1} ,v_{2})$ being collapsed:}

{\small \begin{equation*}
v=\frac{v_1 +v_2}{2}
\end{equation*}}

\begin{lstlisting}
vec3 GetFallbackVertex(IEdge* collapse_edge)
{
  auto verts = collapse_edge->GetVertices();
  return 0.5 * (verts[0]->GetPosition() + verts[1]->GetPosition());
}
\end{lstlisting}

\section{Constraint selection criteria}
\label{sec:constraint-selection}

{\small To solve for the new vertex $v$, a system of linear equations is formed from several constraints as discussed in the earlier sections. Each linear equation is of the form:}

{\small \begin{equation*}
a^T v\ =\ b
\end{equation*}}

{\small Here, each $\displaystyle a$ represents the normal of a constraint plane, and $\displaystyle b$ is the corresponding offset. Geometrically, we are finding the point $\displaystyle v$ that lies at the intersection of all these planes in $\displaystyle \mathbb{R}^{3}$.}

{\small In theory, only three linearly independent constraints (planes) are needed to uniquely determine a point in 3D. However, the algorithm includes more than three constraints to ensure robustness. That’s because:}

\begin{itemize}
\item {\small some constraints may become redundant (linearly dependent).}
\item {\small some may become degenerate in flat or symmetric regions.}
\end{itemize}

{\small The final vertex placement includes only the best three by checking for linear independence and stability using the following criteria while adding constraints one by one:}

\noindent{\small \textbf{First Constraint(}$\displaystyle \mathbf{a_{1}}$\textbf{): Is it valid?}}

{\small \begin{equation*}
a_{1} \neq 0
\end{equation*}}

\begin{lstlisting}
bool IsFirstConstraintValid(const vec3 &A, const mat3 &H, const vec3 &c)
{
    return length(A) > 0;
}
\end{lstlisting}

\noindent {\small \textbf{Second Constraint(}$\displaystyle \mathbf{a_{2}}$\textbf{): Is it linearly independent from the first?}}

{\small \begin{equation*}
( a_{1} \cdotp a_{2})^{2} < ( \| a_{1} \| \ \| a_{2} \| \cos( \alpha ))^{2}
\end{equation*}}
{\small This checks if the angle $\displaystyle \theta $ between the first and second constraint normals is sufficiently large, that is, that they are not almost parallel. $\displaystyle \alpha $ is a threshold angle used to determine acceptable linear independence.}

\begin{lstlisting}
bool IsSecondConstraintValid(const vec3 &A, float alpha, const mat3 &H, const vec3 &c)
{
    float lhs = dot(H[0], A);
    float rhs = length(H[0]) * length(A) * cos(alpha);
    return lhs * lhs < rhs * rhs;  
}
\end{lstlisting}

\noindent {\small \textbf{Third Constraint(}$\displaystyle \mathbf{a_{3}}$\textbf{): Is it not coplanar with the first two?}}
{\small \begin{equation*}
\left(\left( a_{1} \times a_{2}\right) \cdotp a_{3}\right)^{2} < \left( \| a_{1} \times a_{2} \| \ \| a_{3} \| \ \sin( \alpha )\right)^{2}
\end{equation*}}
{\small This ensures that the third constraint’s normal $\displaystyle a_{3}$ does not lie in the plane formed by the normals of the first two constraints, again up to a threshold angle $\displaystyle \alpha $. This guarantees that the three planes intersect at a single point in 3D space, defining a unique solution for the new vertex.}

\begin{figure}[!ht]
\centering
   \includegraphics[width=0.65\columnwidth]{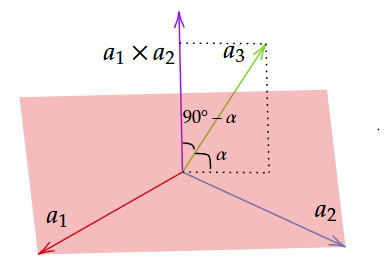}
   \caption{\label{fig:constraint-normal-viz} Visualization of 3 constraint normals}
\end{figure}

{\small Note that, here we use $\displaystyle \sin( \alpha )$ and not $\displaystyle \cos( \alpha )$ when computing the dot product. We define $\displaystyle \alpha $ as the angle between the plane formed by $\displaystyle a_{1}$ and $\displaystyle a_{2}$ and the new vector $\displaystyle a_{3}$. So as \autoref{fig:constraint-normal-viz} suggests, the angle between the vectors $\displaystyle a_{1} \times a_{2}$ and $\displaystyle a_{3}$ will be $\displaystyle 90\degree -\alpha $. So we have,}

{\small \begin{equation*}
\begin{aligned}
\left( a_{1} \times a_{2}\right) \cdotp a_{3} & =\| a_{1} \times a_{2} \| \ \| a_{3} \| \cos( 90\degree -\alpha ) & \\
 & =\| a_{1} \times a_{2} \| \ \| a_{3} \| \sin( \alpha ) & 
\end{aligned}
\end{equation*}}

\begin{lstlisting}
bool IsThirdConstraintValid(const vec3 &A, float alpha, const mat3 &H, const vec3 &c)
{
    vec3 p = cross(H[0], H[1]);

    float lhs = dot(p, A);
    float rhs = length(p) * length(A) * sin(alpha);
    return lhs * lhs < rhs * rhs;
}
\end{lstlisting}

\section{Vertex placement}
\label{sec:vertex-placement}

{\small Putting it all together, we revisit the overall edge collapse algorithm and implement all its components. We begin by computing the constraints described above in an order that best suits our use case. Next, we apply the constraint selection criteria to form a set of solvable constraints. Finally, we compute the optimal vertex position based on this set, resorting to our fallback strategy if the system remains unsolvable. The resulting optimal position and its associated cost are then returned for further handling in the priority queue.}

\begin{lstlisting}
void GetCollapseVertex(const IMesh* mesh, const IEdge &collapse_edge, vec3 &collapse_vertex, float &collapse_error)
{
  vector<IConstraint> constraints;

  IConstraint qem = QEM(mesh, collapse_edge);                     constraints.add(qem);
  IConstraint volume_pres = VolumePres(mesh, collapse_edge);      constraints.add(volume_pres);
  IConstraint volume_opt = VolumeOpt(mesh, collapse_edge);        constraints.add(volume_opt);
  IConstraint boundary_qem = BoundaryQEM(mesh, collapse_edge);    constraints.add(boundary_qem);
  IConstraint boundary_pres = BoundaryPres(mesh, collapse_edge);  constraints.add(boundary_pres);
  IConstraint boundary_opt = BoundaryOpt(mesh, collapse_edge);    constraints.add(boundary_opt);
  IConstraint tri_opt = TriangleOpt(mesh, collapse_edge);         constraints.add(tri_opt);

  int n_eqns = 0;
  float alpha = radians(5.0);  // 5 degrees, in radians
  mat3 H; vec3 c;
  
  for (auto constraint : constraints)
  {
    if (n_eqns == 3) break;

    mat3 constr_H = constraint.GetH();  vec3 constr_c = constraint.GetC();

    for(int i = 0; i < 3; i++)
    {
      vec3 A = constr_H[i]; float b = constr_C[i];

      if (n_eqns == 0 && IsFirstConstraintValid(A, H, c))
      {
          H[0] = A;  c[0] = b;
          n_eqns++;
      }
      else if (n_eqns == 1 && IsSecondConstraintValid(A, alpha, H, c))
      {    
          H[1] = A;  c[1] = b;
          n_eqns++;
      }
      else if (n_eqns == 2 && IsThirdConstraintValid(A, alpha, H, c))
      {
          H[2] = A;  c[2] = b;
          n_eqns++;
          break;
      }
    }
  }
  
  collapse_vertex = determinant(H) == 0.0 ? GetFallbackVertex(collapse_edge) : inverse(H) * c;
  
  collapse_error = 0;
  for (auto constraint : constraints)
  {
    // Note: The `EvaluateCost` function can be weighted here.
    // The weights can be user-defined to prioritize
    // different types of mesh deformation minimization
    // (e.g., prioritizing volume preservation
    // over triangle quality)
    collapse_error += constraint.EvaluateCost(collapse_error);
  }
}
\end{lstlisting}

\section{Handling mesh attributes}

{\small The focus of the paper so far has been solely on the geometry and topology of 3D meshes during simplification. However, real-world meshes often include attributes like per-vertex colors and normals or per-face material indices used in rendering. When simplifying such meshes via edge collapse, it’s crucial to preserve attribute consistency at new vertices to avoid visual artifacts like color seams. The following sections explore key methods addressing this challenge.}

\subsection{Higher-dimensional quadrics for QEM}

{\small \cite{Garland1998Simplifying} presented a modification to the QEM method that incorporates vertex attributes in addition to position when computing the new vertex in an edge collapse. This approach uses higher-dimensional quadrics, in which each vertex is expressed as a vector in a higher-dimensional space $\displaystyle \mathbb{R}^{n}$, $\displaystyle n >3$, combining its position with all associated attributes. For instance, if a vertex at $\displaystyle ( x,y,z)$ has only color attributes $\displaystyle ( R,G,B)$, it is represented as a vector $\displaystyle ( x,y,z,R,G,B) \in \mathbb{R}^{6}$.}

{\small Even when vertex coordinates are given in more than three dimensions, the QEM error measure (sum of distances of the new point from adjacent planes) can still be used. The distance to each plane is now computed using $\displaystyle n$ dimensions but is still with respect to a 2D plane, because three non-collinear points define a plane no matter how many dimensions they are in.}

{\small To compute the distance of a point $\displaystyle v$ from a plane, we first determine a vector $\displaystyle u$ from $\displaystyle v$ to the plane that is perpendicular to its surface. To do this, we define an orthonormal basis in which two basis vectors span the plane and denote these as $\displaystyle e_{1}$ and $\displaystyle e_{2}$. These two vectors are part of a complete orthonormal basis $\{e_{1} ,e_{2} ,\dotsc ,e_{n}\}$ for all of $\displaystyle \mathbb{R}^{n}$. The exact values of $\{e_{3} ,\dotsc ,e_{n}\}$ need not be known. We only need to know that such a basis exists. If needed, they can be determined using the Gram-Schmidt process, which begins with $\displaystyle n$ linearly independent vectors and iteratively removes components parallel to previously computed basis vectors.}

{\small Let the plane be determined by three points $\displaystyle p,q,r\in \mathbb{R}^{n}$.}

\begin{figure}[!ht]
\centering
   \includegraphics[width=0.65\columnwidth]{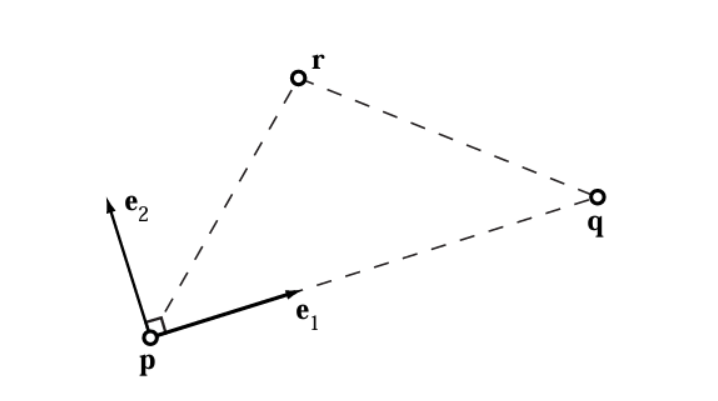}
   \caption{\label{fig:highdim-quadrics-ref-frame} Local frame of reference in $\mathbb{R}^n$}
\end{figure}

{\small As shown in \autoref{fig:highdim-quadrics-ref-frame}, two orthogonal axes $\displaystyle e_{1}$ and $\displaystyle e_{2}$ can be defined on the plane spanned by $\displaystyle ( p,q,r)$:}
\begin{itemize}
\item {\small $\displaystyle e_1$ is the unit vector in the direction $\displaystyle q-p$, i.e., $\displaystyle e_1=\frac{q-p}{\begin{Vmatrix}q-p\end{Vmatrix}}$}
\item {\small Using the Gram-Schmidt process, $\displaystyle e_2$ can be obtained by taking the vector $\displaystyle r-p$, removing its projection onto $\displaystyle e_{1}$ (to ensure orthogonality with $\displaystyle e_{1}$), and then normalizing. In other words,}
\end{itemize}

{\small \begin{equation*}
e_{2} \ =\frac{( r-p) \ -\ ( e_{1} \cdotp ( r-p)) e_{1}}{\begin{Vmatrix}
( r-p) \ -\ ( e_{1} \cdotp ( r-p)) e_{1}
\end{Vmatrix}}
\end{equation*}}

{\small Next, a point $\displaystyle p$ lying on the plane is chosen, and the vector from $\displaystyle p$ to $\displaystyle v$, i.e $\displaystyle w=v-p$, is expressed as the sum of its components along the orthonormal basis $\{e_{1} ,e_{2} ,\dotsc ,e_{n}\}$:}

{\small \begin{equation*}
{\textstyle w=\sum _{i=1}^{n}( w\cdotp e_{i}) e_{i}}
\end{equation*}}

{\small Next, the components along the plane i.e $\displaystyle ( e_{1} ,e_{2})$ are removed from $\displaystyle w$, giving us a vector $\displaystyle u$:}

{\small \begin{equation*}
u=w-( w\cdotp e_{1}) e_{1} -( w\cdotp e_{2}) e_{2} =v-(( w\cdotp e_{1}) e_{1} +( w\cdotp e_{2}) e_{2} +p)
\end{equation*}}

{\small Here, $\displaystyle u$ is the perpendicular from $\displaystyle v$ to the plane, whose squared length is precisely $\displaystyle \Eulerconst $:}

{\small \begin{equation*}
\begin{aligned}
\Eulerconst  & =u\cdotp u & \\
 & =( w-( w\cdotp e_1) e_1 -( w\cdotp e_2) e_2) \cdotp ( w-( w\cdotp e_1) e_1 -( w\cdotp e_2) e_2) & \\
 & =\| w\| ^{2} +( w\cdotp e_{1})^{2}( e_{1} \cdotp e_{1}) +( w\cdotp e_{2})^{2}( e_{2} \cdotp e_{2}) & \dotsc e_{1} \cdotp e_{1} =e_{2} \cdotp e_{2} =1\\
 & \ -2( w\cdotp e_{1})^{2} -2( w\cdotp e_{2})^{2} +\ \bcancel{2( w\cdotp e_{1})( w\cdotp e_{2})( e_{1} \cdotp e_{2})} & \dotsc e_{1} \cdotp e_{2} =0\ \\
 & =\| w\| ^{2} -( w\cdotp e_{1})^{2} -( w\cdotp e_{2})^{2} & 
\end{aligned}
\end{equation*}}
{\small Expanding this out further using $\displaystyle w=v-p$, we get:}

{\small \begin{equation*}
\begin{aligned}
\Eulerconst  & =( v-p) \cdotp ( v-p) -(( v-p) \cdotp e_{1})^{2} -(( v-p) \cdotp e_{2})^{2} \\
 & =( v\cdotp v+p\cdotp p-2p\cdotp v) -( e_{1} \cdotp v-e_{1} \cdotp p)^{2} -( e_{2} \cdotp v-e_{2} \cdotp p)^{2} \\
 & =\left( v\cdotp v-( e_{1} \cdotp v)^{2} -( e_{2} \cdotp v)^{2}\right) +2( -p\cdotp v+( e_{1} \cdotp p)( e_{1} \cdotp v) +( e_{2} \cdotp p)( e_{2} \cdotp v)) +\left( p\cdotp p-( e_{1} \cdotp p)^{2} -( e_{2} \cdotp p)^{2}\right) \\
 & =\left( v^{T} v-v^{T} e_{1} e_{1}^{T} v-v^{T} e_{2} e_{2}^{T} v\right) +2(( p\cdotp e_{1}) e_{1} +( p\cdotp e_{2}) e_{2} -p) \cdotp v+\left( p\cdotp p-( p\cdotp e_{1})^{2} -( p\cdotp e_{2})^{2}\right) \\
 & =v^{T}\textcolor[rgb]{0.82,0.01,0.11}{\left(I-e_1 {e_1}^T-e_2{e_2}^T\right)} v+2
 \textcolor[rgb]{0.29,0.56,0.89}{((p\cdotp e_1)e_1 +(p\cdotp e_2)e_2-p)}^T v+
 \textcolor[rgb]{0.74,0.06,0.88}{\left(p\cdotp p-{(p\cdotp e_1)}^{2}-{(p\cdotp e_2)}^{2}\right)} \\
 & =v^{T}\textcolor[rgb]{0.82,0.01,0.11}{H} v+2\textcolor[rgb]{0.29,0.56,0.89}{c}^{T} v+\textcolor[rgb]{0.74,0.06,0.88}{k}
\end{aligned}
\end{equation*}}

{\small Note that this expression matches the form of the error used in the original QEM method. This means that, aside from differences in the values and dimensions of the entities $\{H,c,k\}$, the procedure for calculating the cost and determining the optimal $\displaystyle v$ remains unchanged: the optimal $\displaystyle v$ is still given by $\displaystyle -H^{-1} c$.}

{\small Furthermore, once computed, $\displaystyle v$ not only represents the optimal vertex position but also encodes the optimal values for all associated scalar attributes.}

\subsection{Energy-based cost computation}

{\small \cite{Hoppe1996Progressive} introduced progressive meshes - sequences of meshes representing varying levels of detail of an input mesh, each created through successive edge collapse operations. They propose an alternative method to compute the cost of each edge collapse by defining it as the difference in an energy function. The cost reflects how much the energy function of the mesh changes before and after the collapse, with edges causing smaller energy differences considered better candidates for collapse. Their approach also handles discrete face attributes and scalar vertex attributes at each level of detail.}

\subsubsection{Prerequisites}

\begin{figure}[!ht]
\centering
   \includegraphics[width=0.65\columnwidth]{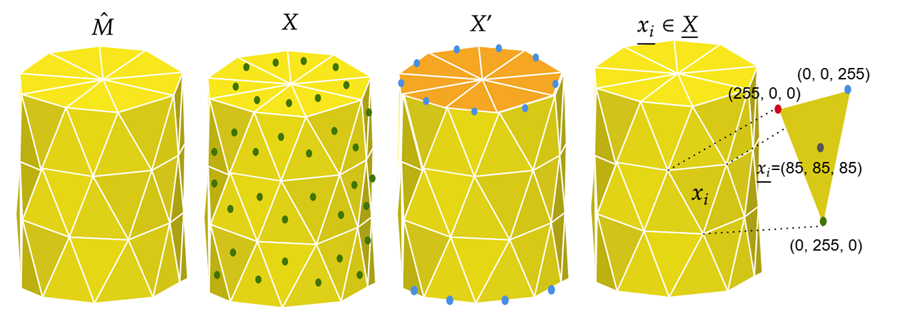}
   \caption{\label{fig:orig-mesh-sampled-pts} An example showing the original mesh and sampled point sets }
\end{figure}

{\small Before introducing the cost function, we define key geometric entities and setup steps needed to compute the cost of an edge collapse. \autoref{fig:orig-mesh-sampled-pts} shows an example.}

\noindent {\small \textbf{\underline{Original mesh definition:}}}

{\small The original mesh (before any edge collapse operations) is denoted as:}

\begin{center}
{\small $\displaystyle \hat{M} =\left(\hat{V} ,\hat{F} ,\hat{\underline{V}} ,\hat{\underline{F}}\right)$ }
\end{center}

\begin{itemize}
\item {\small \textbf{Vertices:} $\displaystyle \hat{V} =\{v_{1} ,\dotsc ,v_{n}\}$ , where each $\displaystyle v_{i} \in \mathbb{R}^{3}$}
\item {\small \textbf{Faces:} $\displaystyle \hat{F} =\{f_{1} ,\dotsc ,f_{m}\}$, with each $\displaystyle f_{i} \in \{1,\dotsc ,n\}^{3}$}
\item {\small \textbf{Vertex} \textbf{attributes:} $\displaystyle \hat{\underline{V}} =\left\{\underline{v_{1}} ,\dotsc ,\underline{v_{n}}\right\}$, with each $\displaystyle \underline{v_{i}} \in \mathbb{R}^{d}$ (for example, $\displaystyle d=3$ for RGB colors)}
\item {\small \textbf{Face} \textbf{attributes:} $\displaystyle \underline{\hat{F}} =\left\{\underline{f_{1}} ,\dotsc ,\underline{f_{m}}\right\}$ i, with each $\displaystyle \underline{f_{i}} \in \mathbb{Z}^{d'}$ (for example, $\displaystyle d'=1$ for material indices)}

\end{itemize}

\noindent\textbf{{\small \underline{Setup for simplification:}}}

{\small Before simplification, the following steps are performed:}

\begin{enumerate}
\item {\small \textbf{Surface sampling: }}

{\small Sample a set $\displaystyle X$ of $\displaystyle k$ points on the surface of $\displaystyle \hat{M}$:}

{\small $\displaystyle X=\{x_{1} ,\dotsc ,x_{k}\} ,\ x_{i} \in \mathbb{R}^{3}$}

{\small These points will serve to approximate $\displaystyle \hat{M}$ in the energy functions.}
\item {\small \textbf{Attribute sampling:}}

{\small For each $\displaystyle x_{i}$, compute its scalar attribute $\displaystyle \underline{x_{i}} \in \mathbb{R}^{d}$ via barycentric interpolation on its containing face yielding the attribute set $\displaystyle \underline{X}$ \ corresponds to $\displaystyle X$:}

{\small $\displaystyle \underline{X} =\left\{\underline{x_{1}} ,\dotsc ,\underline{x_{k}}\right\} ,\ \underline{x_{i}} \in \mathbb{R}^{d}$}
\item {\small \textbf{Sharp-edge sampling:}}

{\small Sample an additional set $\displaystyle X'$ of $\displaystyle k'$ points constrained to lie only on \textit{sharp edges }on the surface of $\displaystyle \hat{M}$:}

{\small $\displaystyle X'=\{x'_{1} ,\dotsc ,x'_{k'}\} ,\ x'_{i} \in \mathbb{R}^{3}$}

{\small \textit{Sharp edges }are comprised of:}
\begin{enumerate}
\item {\small Material boundaries: edges between faces with $\displaystyle \underline{f_{i}} \neq \underline{f_{j}}$. }
\item {\small Geometric boundaries: Edges connected to only one face.}
\end{enumerate}
\end{enumerate}

\subsubsection{Cost function definitions}

{\small Consider the mesh $\displaystyle \hat{M}$ after some edge collapses. Let $\displaystyle M^{\uparrow }$ be the state before collapsing another edge, and $\displaystyle M$ the state after edge collapse. The cost for this collapse is formulated as:}

{\small \begin{equation} \label{eq:energy-cost-fn}
\Eulerconst =\underbrace{\left( E_{dist}( M) -E_{dist}\left( M^{\uparrow }\right)\right)}_{\Delta E_{dist}} +\underbrace{\left( E_{spring}( M) -E_{spring}\left( M^{\uparrow }\right)\right)}_{\Delta E_{spring}} +\underbrace{\left( E_{scalar}( M) -E_{scalar}\left( M^{\uparrow }\right)\right)}_{\Delta E_{scalar}} +D_{disc}
\end{equation}}

\noindent\textbf{{\small \underline{Distance energy:}}}

{\small $\displaystyle E_{dist}$ measures geometric fidelity by penalizing the distance between each sampled point $\displaystyle x_i \in X$ and its projection $\displaystyle p( x_i)$ on \ the mesh $\displaystyle M$.}

{\small \begin{equation}
E_{dist}( M) =\sum _{x_i \in X}\Vert x_i -p( x_i)\Vert ^{2} \tag{4a}
\end{equation}}

\noindent{\small \textbf{\underline{Spring energy:}}}

{\small $\displaystyle E_{spring}$ regularizes the optimization by treating each edge as a "spring", contributing its squared length to the energy.}

{\small \begin{equation}
E_{spring}( M) =\kappa \sum _{( v_{a} ,v_{b}) \ \in \ \text{edges}( M)}\Vert v_{a} -v_{b}\Vert ^{2} \tag{4b}
\end{equation}}

{\small Here, the spring constant $\displaystyle \kappa $ weights this term.}\newline

\noindent{\small \textbf{\underline{Scalar attribute energy:}}}

{\small $\displaystyle E_{scalar}$ is similar to $\displaystyle E_{dist}$, but for scalar attributes. It measures the squared difference between the attributes of each sampled point $\displaystyle \underline{x_{i}}$ and those of its projection $\displaystyle \underline{p}( x_{i})$:}

{\small \begin{equation}
E_{scalar}( M) =c_{scalar}\sum _{\underline{x_{i}} \in \underline{X}}\left\Vert \underline{x_{i}} -\underline{p}( x_{i})\right\Vert ^{2} \tag{4c}
\end{equation}}
{\small where $\displaystyle c_{scalar}$ weights this term.}\newline

\noindent{\small \textbf{\underline{Discontinuity preservation offset:}}}

{\small $\displaystyle D_{disc}$ penalizes collapsing a \textit{sharp edge} $\displaystyle e$ that affects discontinuities tracked by $\displaystyle X'$:}

{\small \begin{equation}
D_{disc} =\text{numProject}( X',e) \ \Vert e\Vert ^{2} \tag{4d}
\end{equation}}

{\small Here, $\displaystyle \text{numProject}( X',e)$ denotes the number of points in $\displaystyle X'$ projecting onto $\displaystyle e$. This term is applied only if the collapse alters the discontinuity connectivity, based on criteria presented in \cite{Hoppe1996Progressive}. To forbid the collapses entirely, set $\displaystyle D_{disc} =\infty $.}\newline

\noindent{\small \textbf{\underline{Cost function evaluation and optimal vertex placement:}}}

{\small Recall that $\displaystyle M^{\uparrow }$ is the state of the mesh before collapsing the current edge, and $\displaystyle M$ the mesh after the collapse. The new vertex is placed at position $\displaystyle v$ with attributes $\displaystyle \underline{v}$. We need to compute $\displaystyle v$ and $\displaystyle \underline{v}$ that minimize the cost $\displaystyle \Eulerconst $.}

{\small As per \autoref{eq:energy-cost-fn}, the terms in $\displaystyle \Eulerconst $ depend on $\displaystyle v$ and $\displaystyle \underline{v}$ as follows:}
\begin{itemize}
\item {\small $\displaystyle E_{dist}( M)$ and $\displaystyle E_{spring}( M)$ depend only on $\displaystyle v$.}
\item {\small $\displaystyle E_{scalar}( M)$ depends only on $\displaystyle \underline{v}$.}
\item {\small $\displaystyle D_{disc}$ and the energy terms for $\displaystyle M^{\uparrow }$ are independent of both $\displaystyle v$ or $\displaystyle \underline{v}$. So, they only affect the cost $\displaystyle \Eulerconst $ but not the optimization.}
\end{itemize}

{\small Thus, here are the steps to compute $\displaystyle \Eulerconst $, $\displaystyle v$ and $\displaystyle \underline{v}$ for a given edge collapse:}
\begin{enumerate}
\item {\small Minimize $\displaystyle \Delta E_{dist} +\Delta E_{spring}$ over $\displaystyle v$.}
\item {\small Minimize $\displaystyle \Delta E_{scalar}$ over $\displaystyle \underline{v}$.}
\item {\small If the discontinuity criteria hold, compute $\displaystyle D_{disc}$.}
\item {\small Return $\displaystyle \Eulerconst =\Delta E_{dist} +\Delta E_{spring} +\Delta E_{scalar} +D_{disc}$ along with the optimal $\displaystyle v$ and $\displaystyle \underline{v}$.}
\end{enumerate}

\subsubsection{Optimization process}

\begin{figure}[!ht]
\centering
   \includegraphics[width=0.65\columnwidth]{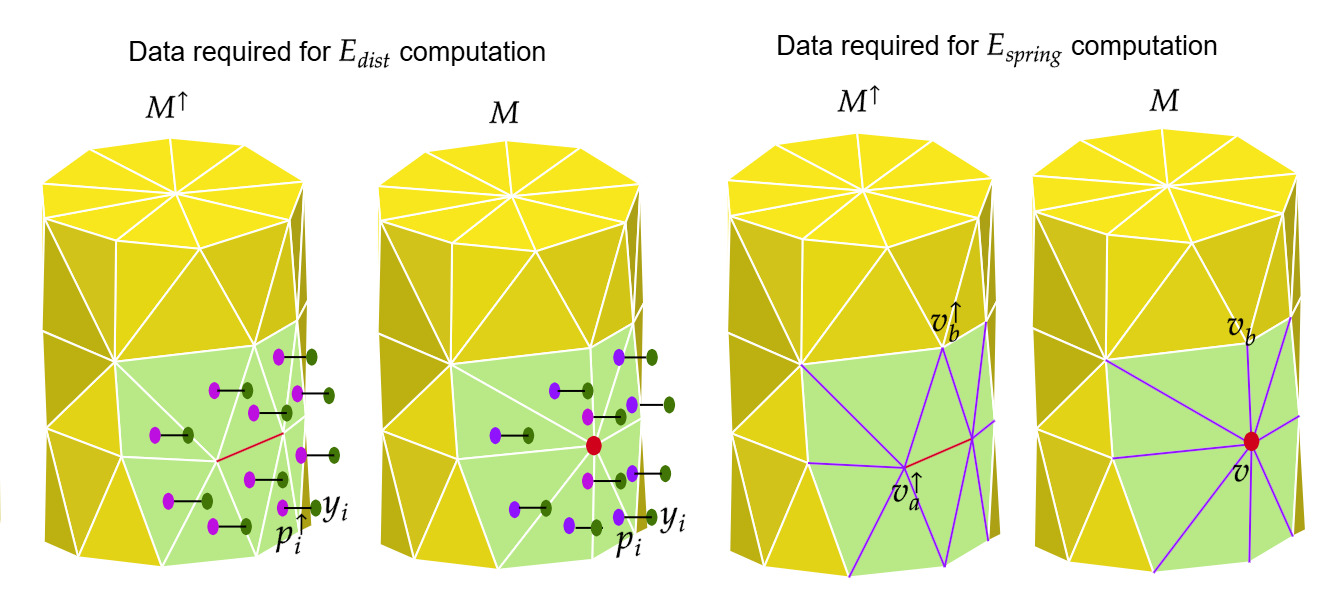}
   \caption{\label{fig:data-for-dist-spring} 
An example showing data required for $\displaystyle E_{dist}$ and $\displaystyle E_{spring}$}
\end{figure}

{\small Minimizing the error $\displaystyle \Eulerconst $ is more complex than previous error functions and requires a staged optimization process. We describe that process in detail below, and the data required for computing and optimizing the error is shown in an example in \autoref{fig:data-for-dist-spring}.}

\noindent{\small \textbf{\underline{Minimizing }}$\displaystyle \Delta E_{spring}$\underline{\textbf{ over }}$\displaystyle v$}

{\small \begin{equation*}
\begin{aligned}
\Delta E_{spring} & =E_{spring}( M) -E_{spring}\left( M^{\uparrow }\right)\\
 & =\kappa \left[\sum _{( v_{a} ,v_{b}) \ \in \ \text{edges}( M)}\Vert v_{a} -v_{b}\Vert ^{2} -\sum _{\left( v_{a}^{\uparrow } ,v_{b}^{\uparrow }\right) \ \in \ \text{edges}\left( M^{\uparrow }\right)}\left\Vert v_{a}^{\uparrow } -v_{b}^{\uparrow }\right\Vert ^{2}\right]
\end{aligned}
\end{equation*}}
{\small We can observe that any edge with unchanged endpoints in $\displaystyle M$ and $\displaystyle M^{\uparrow }$ cancels out in $\displaystyle \Delta E_{spring}$. Since differences occur only near the collapsed edge, only two edge sets contribute nonzero terms:}
\begin{enumerate}
\item {\small $\displaystyle N$: \ edges in $\displaystyle M$ incident on the new vertex at $\displaystyle v$.}
\item {\small $\displaystyle N^{\uparrow }$: edges in $\displaystyle M^{\uparrow }$ incident on the collapsed edge.}
\end{enumerate}

{\small Thus, we can simplify $\displaystyle \Delta E_{spring}$ as follows:}
{\small \begin{equation}
\Delta E_{spring} =\kappa \left[\sum _{( v ,v_{b}) \ \in \ N}\Vert v-v_{b}\Vert ^{2} -\sum _{\left( v_{a}^{\uparrow } ,v_{b}^{\uparrow }\right) \ \in \ N^{\uparrow }}\left\Vert v_{a}^{\uparrow } -v_{b}^{\uparrow }\right\Vert ^{2}\right]
\end{equation}}
{\small where $\displaystyle v$ appears explicitly as all edges in $\displaystyle N$ are incident on it.}

{\small Since $\displaystyle E_{spring}\left( M^{\uparrow }\right)$ is independent of $\displaystyle v$, it is ignored in the optimization process and added only to the final cost. }

{\small The spring term $\displaystyle E_{spring}( M)$,}
{\small \begin{equation*}
E_{spring}( M) =\kappa \sum _{( v ,v_{b}) \ \in \ N}\Vert v-v_{b}\Vert ^{2}
\end{equation*}}
{\small is a quadratic in $\displaystyle v$ and can be written as}
{\small \begin{equation*}
\begin{aligned}
E_{spring}( M) & =\kappa \sum _{( v ,v_{b}) \ \in \ N}\left( v^{T} v-2v_{b}^{T} v+v_{b}^{T} v_{b}\right) & \\
 & =\kappa \left(\sum _{( v ,v_{b}) \ \in \ N} v^{T} v-\sum _{( v ,v_{b}) \ \in \ N} 2v_{b}^{T} v+\sum _{( v ,v_{b}) \ \in \ N} v_{b}^{T} v_{b}\right) & \\
 & =\kappa \left( v^{T}(\textcolor[rgb]{0.82,0.01,0.11}{n}\textcolor[rgb]{0.82,0.01,0.11}{_{e}}\textcolor[rgb]{0.82,0.01,0.11}{I}) v-2\left(\textcolor[rgb]{0.29,0.56,0.89}{\sum _{( v ,v_{b}) \ \in \ N}}\textcolor[rgb]{0.29,0.56,0.89}{v}\textcolor[rgb]{0.29,0.56,0.89}{_{b}}\right)^{T} v+\left(\textcolor[rgb]{0.74,0.06,0.88}{\sum _{( v ,v_{b}) \ \in \ N}}\textcolor[rgb]{0.74,0.06,0.88}{v}\textcolor[rgb]{0.74,0.06,0.88}{_{b}^{T}}\textcolor[rgb]{0.74,0.06,0.88}{v}\textcolor[rgb]{0.74,0.06,0.88}{_{b}}\right)\right) & \dotsc \ n_{e}\text{ is the number of edges in } N\\
 & =\kappa \left( v^{T}\textcolor[rgb]{0.82,0.01,0.11}{H} v+2\textcolor[rgb]{0.29,0.56,0.89}{c}^{T} v+\textcolor[rgb]{0.74,0.06,0.88}{k}\right) & 
\end{aligned}
\end{equation*}}
 
\noindent {\small \underline{\textbf{Minimizing} }$\displaystyle \Delta E_{dist}$\underline{\textbf{ over }}$\displaystyle v$}

{\small \begin{equation*}
\begin{aligned}
\Delta E_{dist} & =E_{dist}( M) -E_{dist}\left( M^{\uparrow }\right)\\
 & =\sum _{x_{i} \in X}\left[\Vert x_{i} -p( x_{i})\Vert ^{2} -\left\Vert x_{i} -p^{\uparrow }( x_{i})\right\Vert ^{2}\right]
\end{aligned}
\end{equation*}}
{\small where the projection of the same sampled point $\displaystyle x_{i}$ is $\displaystyle p( x_{i})$ in $\displaystyle M$ and $\displaystyle p^{\uparrow }( x_{i})$ in $\displaystyle M^{\uparrow }$.}

{\small Since $\displaystyle M$ and $\displaystyle M^{\uparrow }$ differ only in the vicinity of the one edge being collapsed, most points project identically in both states and cancel out. Thus, only points $\displaystyle Y\subseteq X$ projecting onto the neighborhood of the edge contribute:}

{\small \begin{equation}
\Delta E_{dist} =\sum _{y_{i} \in Y}\left[\Vert y_{i} -p( y_{i})\Vert ^{2} -\left\Vert y_{i} -p^{\uparrow }( y_{i})\right\Vert ^{2}\right]
\end{equation}}

{\small We define the neighborhood of an edge as the set of faces connected to either endpoint of that edge in $\displaystyle M$ or $\displaystyle M^{\uparrow }$.}

{\small This simplification is based on a \textit{locality assumption}: the new vertex $\displaystyle v$ stays near the collapsed edge, so the projections of distant points in $\displaystyle X$ remain unchanged. Although allowing $\displaystyle v$ farther away could lower the cost, it would require recomputing over all $\displaystyle X$. In practice, restricting $\displaystyle v$ and using the simplified $\displaystyle \Delta E_{dist}$ works well.}

{\small Since $\displaystyle E_{dist}\left( M^{\uparrow }\right)$ is independent of $\displaystyle v$, it is ignored in the optimization process and added only to the final cost. }

{\small Now, although the $\displaystyle E_{dist}( M)$ term,}
{\small \begin{equation*}
E_{dist}( M) =\sum _{y_{i} \in Y}\Vert y_{i} -p( y_{i})\Vert ^{2}
\end{equation*}}
{\small appears independent of $\displaystyle v$, each projection $\displaystyle p( y_{i})$ depends on $\displaystyle v$.}

{\small For a point $\displaystyle y$, the projection $\displaystyle p( y)$ is}
{\small \begin{equation*}
p( y) =\arg\min_{p}\Vert y -p\Vert ^{2}
\end{equation*}}
{\small where $\displaystyle p$ lies on some face in $\displaystyle M$ with vertices $\displaystyle ( v_{a} ,v_{b} ,v_{c})$. We use barycentric coordinates $\displaystyle \beta =( \beta _{a} ,\beta _{b} ,\beta _{c})$ to express $\displaystyle p$:}
{\small \begin{equation*}
\begin{aligned}
p=V\beta , & \ \ \ \ \ \ \ \ V=\begin{bmatrix}
v_{a} & v_{b} & v_{c}
\end{bmatrix} \in \mathbb{R}^{3\times 3} ,\  & \ \ \ \beta \in [ 0,1]^{3}
\end{aligned}
\end{equation*}}
{\small We can define a function $\displaystyle \beta ( y)$ to denote the projected barycentric coordinates for any point $\displaystyle y$:}
{\small \begin{equation*}
\begin{aligned}
\beta ( y) =\arg\min_{\beta }\Vert y-V\beta \Vert ^{2} ,\ \ \  & \ \ \ \ \ \ \ \ p( y) =V\beta ( y)
\end{aligned}
\end{equation*}}

{\small Now, since both $\displaystyle V$ and $\displaystyle \beta $ are face-specific, we extend $\displaystyle V$ to the full mesh $\displaystyle M$ with $\displaystyle n$ vertices $\displaystyle \{v_{1} ,\dotsc ,v,\dotsc ,v_{n}\}$ and any given $\displaystyle \beta $ to an $\displaystyle n$-dimensional vector, zeroing out the entries for all vertices except $\displaystyle ( v_{a} ,v_{b} ,v_{c})$}
{\small \begin{equation*}
\begin{aligned}
\mathbf{V} =[ v_{1} \mid \dotsc \mid v_{n}] \in \mathbb{R}^{3\times n} , & \ \ \ \ \ \ \ \ \mathbf{\upbeta } \in [ 0,1]^{n}
\end{aligned}
\end{equation*}}
{\small Thus, $\displaystyle E_{dist}( M)$ can be rewritten in terms of $\displaystyle v$ (as contained within $\displaystyle \mathbf{V}$):}
{\small \begin{equation*}
E_{dist}( M) =\sum _{y_{i} \in Y}\Vert y_{i} -p( y_{i})\Vert ^{2} \Longrightarrow \sum _{y_{i} \in Y}\Vert y_{i} -\mathbf{V\upbeta }( y_{i})\Vert ^{2} \Longrightarrow \sum _{y_{i} \in Y}\min_{\mathbf{\upbeta }}\Vert y_{i} -\mathbf{V\upbeta }\Vert ^{2}
\end{equation*}}

{\small We can now minimize $\displaystyle E_{dist}( M)$ over $\displaystyle v$. Since evaluating it requires projecting each $\displaystyle y_{i}$ via an inner minimization in $\displaystyle \mathbf{\upbeta }$ space, the problem becomes nested: an outer minimization over $\displaystyle v$ and inner minimizations over all $\displaystyle \mathbf{\upbeta }( y_{i})$.}

\begin{figure}[!ht]
\centering
   \includegraphics[width=0.65\columnwidth]{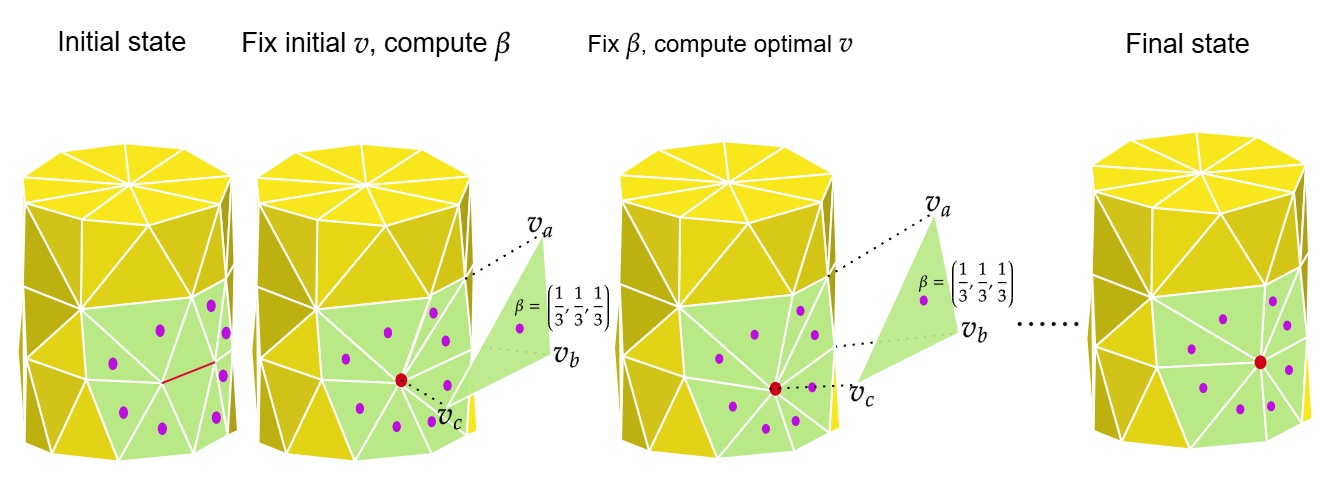}
   \caption{\label{fig:alternating-optimization} 
Alternating optimization of $\displaystyle v$ and each $\displaystyle \mathbf{\upbeta }( y_{i})$}
\end{figure}

{\small We solve the nested minimization iteratively as shown in \autoref{fig:alternating-optimization}, starting with an initial guess for $\displaystyle v$ and alternating between: optimizing $\displaystyle v$ with fixed $\displaystyle \mathbf{\upbeta }( y_{i})$, then updating $\displaystyle \mathbf{\upbeta }( y_{i})$ with fixed $\displaystyle v$. This repeats until convergence, i.e., when the values of $\displaystyle v$ and each $\displaystyle \mathbf{\upbeta }( y_{i})$ don't change much between iterations. In practice, a small number of iterations is sufficient.}

{\small The inner minimization - over each $\displaystyle \mathbf{\upbeta }( y_{i})$ with fixed $\displaystyle v$ - is called the \textit{projection subproblem,} i.e., projecting all points in $\displaystyle Y$ onto $\displaystyle M$. A brute-force method is to try projecting every $\displaystyle y_{i}$ on every face of $\displaystyle M$ and compute $\displaystyle \mathbf{\upbeta }( y_{i})$ corresponding to the closest face. But, \cite{Hoppe1996Progressive} adds two speedups to this approach:}

\begin{enumerate}
\item {\small Use a spatial partitioning structure to find candidate faces in $\displaystyle O( 1)$ time per point, especially useful early on or after edge collapses in new regions.}
\item {\small If $\displaystyle y_{i}$ was projected onto $\displaystyle M^{\uparrow }$, limit its projection on $\displaystyle M$ to the faces neighboring the previous one, leveraging locality.}
\end{enumerate}

{\small The outer minimization - over $\displaystyle v$ while keeping all $\displaystyle \mathbf{\upbeta }( y_{i})$ constant - is now solved by rewriting $\displaystyle E_{dist}( M)$ using that constancy.}

{\small \begin{equation*}
\begin{aligned}
E_{dist}( M) & =\sum _{y_{i} \in Y}\Vert y_{i} -\mathbf{V\upbeta }( y_{i})\Vert ^{2}\\
 & =\sum _{y_{i} \in Y}\left\Vert y_{i} -\begin{pmatrix}
v_{1} & \dotsc  & v & \dotsc  & v_{n}
\end{pmatrix}\begin{pmatrix}
\mathbf{\upbeta }_{1}( y_{i}) & \dotsc  & \mathbf{\upbeta }_{v}( y_{i}) & \dotsc  & \mathbf{\upbeta }_{n}( y_{i})
\end{pmatrix}^{T}\right\Vert ^{2}
\end{aligned}
\end{equation*}}
{\small where $\displaystyle \mathbf{\upbeta }_{v}( y_{i})$ represents $\displaystyle v's$ component in $\displaystyle \mathbf{\upbeta }( y_{i})$. So we have,}
{\small \begin{equation*}
\begin{aligned}
E_{dist}( M) & =\sum _{y_{i} \in Y}\left\Vert y_{i} -\left(\mathbf{\upbeta }_{v}( y_{i}) v+\sum _{v_{j} \neq v}\mathbf{\upbeta }_{j}( y_{i}) v_{j}\right)\right\Vert ^{2} & \\
 & =\sum\limits _{y_{i} \in Y}\Vert z_{i} -\mathbf{\upbeta }_{v}( y_{i}) v\Vert ^{2} & \text{where } z_{i} =y_{i} -\sum _{v_{j} \neq v}\mathbf{\upbeta }_{j}( y_{i}) v_{j}\\
 & =\sum\limits _{y_{i} \in Y}\left(\mathbf{\upbeta }_{v}^{2}( y_{i}) v^{T} v-2\mathbf{\upbeta }_{v}( y_{i}) z_{i}^{T} v+z_{i}^{T} z_{i}\right) & \\
 & = v^{T} \textcolor[rgb]{0.82,0.01,0.11}{ \left(\sum\limits _{y_i \in Y}\mathbf{\upbeta}_v^2(y_i)\right) I } v + 2 \textcolor[rgb]{0.29,0.56,0.89}{ \left(-\sum\limits _{y_i \in Y}\mathbf{\upbeta}_v(y_i)z_i\right) }^T v + \textcolor[rgb]{0.74,0.06,0.88}{ \left(\sum\limits _{y_i \in Y}z_i^Tz_i\right) } & \\
 & =v^{T}\textcolor[rgb]{0.82,0.01,0.11}{H} v+2\textcolor[rgb]{0.29,0.56,0.89}{c}^{T} v+\textcolor[rgb]{0.74,0.06,0.88}{k} & 
\end{aligned}
\end{equation*}}
{\small This yields the same quadratic error form as for $\displaystyle E_{spring}( M)$, allowing us to jointly optimize both by summing their $\displaystyle H,c,k$ coefficients and solving for $\displaystyle v$.}

\noindent{\small \textbf{\underline{Minimizing }}$\displaystyle \Delta E_{scalar}$\textbf{\underline{ over }}$\displaystyle \underline{v}$}

{\small The cost function $\displaystyle \Delta E_{scalar}$ is analogous to $\displaystyle E_{dist}$ and shares its locality assumption, giving the expression}
{\small \begin{equation} \label{eq:scalar-energy-cost}
\begin{aligned}
\Delta E_{scalar} & =E_{scalar}( M) -E_{scalar}\left( M^{\uparrow }\right)\\
 & =c_{scalar}\sum _{\underline{y_{i}} \in \underline{Y}}\left[\left\Vert \underline{y_{i}} -p\left(\underline{y_{i}}\right)\right\Vert ^{2} -\left\Vert \underline{y_{i}} -p^{\uparrow }\left(\underline{y_{i}}\right)\right\Vert ^{2}\right]
\end{aligned}
\end{equation}}

{\small Here, $\displaystyle \underline{Y} \subseteq \underline{X}$ is a local subset of sample attribute vectors. We define this neighborhood as the set of all sample points on the faces adjacent to the edge being collapsed, consistent with the approach used for $\displaystyle E_{dist}$.}

{\small Since $\displaystyle E_{scalar}\left( M^{\uparrow }\right)$ is independent of $\displaystyle \underline{v}$, it is ignored in the optimization process and added only to the final cost. }

{\small The optimization is simplified by reusing the barycentric coordinate sets computation $\displaystyle \upbeta $ and $\displaystyle \upbeta ^{\uparrow }$ from the previous optimization of $\displaystyle \Delta E_{dist}$.}

{\small $\displaystyle p\left(\underline{y_{i}}\right)$ is the attribute vector corresponding to the projected point. Its value can be computed by }
{\small \begin{equation*}
p( y_{i}) =\underline{\mathbf{V}}\mathbf{\upbeta }\left(\underline{y_{i}}\right)
\end{equation*}}
{\small where $\displaystyle \underline{\mathbf{V}}$ is the stacked matrix of vertex attributes of $\displaystyle M$, that is, $\displaystyle \underline{\mathbf{V}} =\begin{pmatrix}
\underline{v_{1}} & \dotsc  & \underline{v} & \dotsc  & \underline{v_{m}}
\end{pmatrix} \in \mathbb{R}^{d\times m}$.}

{\small So, \autoref{eq:scalar-energy-cost} simplifies to:}
{\small \begin{equation*}
\begin{aligned}
E_{scalar}( M) & =\sum _{\underline{y_{i}} \in \underline{Y}}\left\Vert \underline{y_{i}} -p\left(\underline{y_{i}}\right)\right\Vert ^{2}\\
 & =\sum _{\underline{y_{i}} \in \underline{Y}}\left\Vert \underline{y_{i}} -\underline{\mathbf{V}}\mathbf{\upbeta }\left(\underline{y_{i}}\right)\right\Vert ^{2}\\
 & =\sum _{\underline{y_{i}} \in \underline{Y}}\left\Vert \underline{y_{i}} -\begin{pmatrix}
\underline{v_{1}} & \dotsc  & \underline{v} & \dotsc  & \underline{v_{n}}
\end{pmatrix}\begin{pmatrix}
\mathbf{\upbeta }_{1}\left(\underline{y_{i}}\right) & \dotsc  & \mathbf{\upbeta }_{v}\left(\underline{y_{i}}\right) & \dotsc  & \mathbf{\upbeta }_{n}\left(\underline{y_{i}}\right)
\end{pmatrix}^{T}\right\Vert ^{2}
\end{aligned}
\end{equation*}}
{\small where $\displaystyle \mathbf{\upbeta }_{v}\left(\underline{y_{i}}\right)$ represents $\displaystyle v's$ component in $\displaystyle \mathbf{\upbeta }\left(\underline{y_{i}}\right)$}
{\small \begin{equation*}
\begin{aligned}
E_{scalar}( M) & =\sum _{\underline{y_{i}} \in \underline{Y}}\left\Vert \underline{y_{i}} -\left(\mathbf{\upbeta }_{v}\left(\underline{y_{i}}\right)\underline{v} +\sum _{\underline{v_{j}} \neq \underline{v}}\mathbf{\upbeta }_{j}\left(\underline{y_{i}}\right)\underline{v_{j}}\right)\right\Vert ^{2} & \\
 & =\sum\limits _{\underline{y_{i}} \in \underline{Y}}\left\Vert \underline{z_{i}} -\mathbf{\upbeta }_{v}\left(\underline{y_{i}}\right)\underline{v}\right\Vert ^{2} & \text{where } z_{i} =\underline{y_{i}} -\sum _{v_{j} \neq v}\mathbf{\upbeta }_{j}\left(\underline{y_{i}}\right)\underline{v_{j}}\\
 & =\sum\limits _{\underline{y_{i}} \in \underline{Y}}\left(\mathbf{\upbeta }_{v}^{2}\left(\underline{y_{i}}\right)\underline{v}^{T}\underline{v} -2\mathbf{\upbeta }_{v}\left(\underline{y_{i}}\right)\underline{z_{i}^{T}}\underline{v} +\underline{z_{i}}^{T}\underline{z_{i}}\right) & \\
 & = \underline{v}^T \textcolor[rgb]{0.82,0.01,0.11}{ \left(\sum\limits _{y_i \in Y}\mathbf{\upbeta}_v^2\left(\underline{y_i}\right)\right)I } \underline{v} + 2 \textcolor[rgb]{0.29,0.56,0.89}{ \left(-\sum\limits _{y_i \in Y}\mathbf{\upbeta}_v\left(\underline{y_i}\right){\underline{z_i}}\right)^T } \underline{v} + \textcolor[rgb]{0.74,0.06,0.88}{ \left(\sum\limits _{y_i \in Y}\underline{z_i^Tz_i}\right) } & \\
 & =\underline{v}^T\textcolor[rgb]{0.82,0.01,0.11}{H} \underline{v}+2\textcolor[rgb]{0.29,0.56,0.89}{c}^T \underline{v}+\textcolor[rgb]{0.74,0.06,0.88}{k} & 
\end{aligned}
\end{equation*}}
{\small which can be solved in the same way as the other constraints we have seen before.}

\section{Conclusion}

{\small In this paper, we investigated mesh simplification using edge collapse in detail by performing a deep dive into four important papers providing variations on this algorithm: \cite{Garland1997Surface, Lindstrom1998Turk, Garland1998Simplifying, Hoppe1996Progressive}.}

{\small We started by discussing the basics of mesh simplification and the different categories of algorithms that are used for that purpose. We then focused on edge collapse and the half-edge data structure typically used to implement it. Next, we outlined the general algorithm used for simplification via edge collapses, including important edge cases that need to be considered. We then performed an elaborate analysis of the process of computing the error introduced by a candidate vertex placement through a variety of metrics, and discussed how the associated constraints can be assembled to form a solvable system of linear equations that yield the final, optimal vertex placement for a given edge collapse. In the process, we also dealt with other important considerations while performing mesh simplification, such as handling boundary edges and vertex/face attributes.}

{\small We believe this work can help people interested in geometry processing and mesh simplification to understand these potent algorithms and metrics in depth, implement them for their use cases, and inspire further work in this field.}

\small
\bibliographystyle{jcgt}
\bibliography{paper}

\newpage

\section*{Appendix}
\label{sec:appendix}

\noindent \textbf{{\small Explanation: Both planar and non-planar boundaries yield the same formula for the error in boundary preservation.}}

\begin{figure}[!ht]
\centering
   \includegraphics[width=0.65\columnwidth]{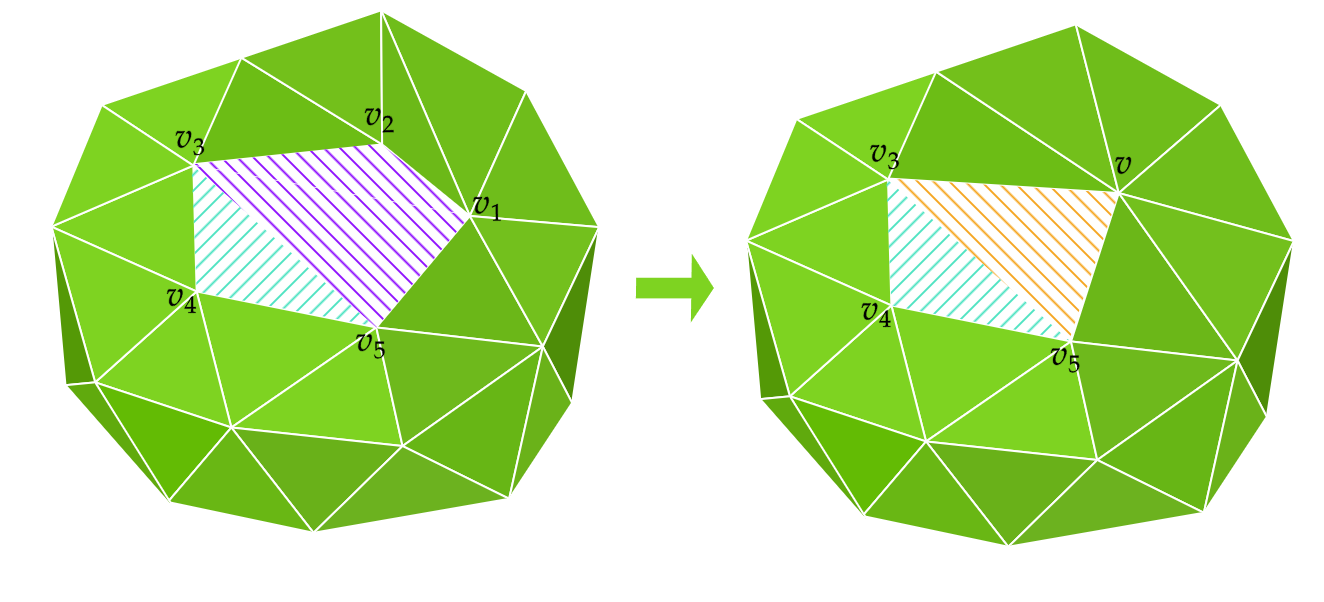}
   \caption{\label{fig:boundary-pres-non-planar} Boundary preservation - non-planar boundary case}
\end{figure}

{\small When the edge $\displaystyle ( v_{1} ,v_{2})$ is replaced by vertex $\displaystyle v$ as per \autoref{fig:boundary-pres-non-planar}, $\displaystyle Area( v_{1} ,\ v_{2} ,\ v_{3} ,v_{5})$ changes but $\displaystyle Area( v_{5} ,\ v_{3} ,\ v_{4})$ remains unchanged. So the change in area in this case, $\displaystyle \Eulerconst '$, will be:}

{\small \begin{equation*}
\begin{aligned}
\Eulerconst ' &= Area( v_{1} ,v_{2} ,v_{3} ,v_{5}) -Area( v,v_{3} ,v_{5}) \ \\
&=Area( v_{1} ,v_{2} ,v_{3}) +Area( v_{1} ,v_{3} ,v_{5}) -Area( v,v_{3} ,v_{5})\\
&=\frac{1}{2}\left(\begin{aligned}
\left( v_{1} \times v_{2}\right) +\left( v_{2} \times v_{3}\right) +\cancel{(\textcolor[rgb]{0.55,0.34,0.16}{v}\textcolor[rgb]{0.55,0.34,0.16}{_{3}}\textcolor[rgb]{0.55,0.34,0.16}{\times v}\textcolor[rgb]{0.55,0.34,0.16}{_{1}}}) & +\\
\left(\cancel{\textcolor[rgb]{0.55,0.34,0.16}{v}\textcolor[rgb]{0.55,0.34,0.16}{_{1}}\textcolor[rgb]{0.55,0.34,0.16}{\times v}\textcolor[rgb]{0.55,0.34,0.16}{_{3}}}\right) +\left(\cancel{\textcolor[rgb]{0.96,0.65,0.14}{v}\textcolor[rgb]{0.96,0.65,0.14}{_{3}}\textcolor[rgb]{0.96,0.65,0.14}{\times v}\textcolor[rgb]{0.96,0.65,0.14}{_{5}}}\right) +\left( v_{5} \times v_{1}\right) & +\\
\left( v_{3} \times v\right) +\left(\cancel{\textcolor[rgb]{0.96,0.65,0.14}{v}\textcolor[rgb]{0.96,0.65,0.14}{_{5}}\textcolor[rgb]{0.96,0.65,0.14}{\times v}\textcolor[rgb]{0.96,0.65,0.14}{_{3}}}\right) +\left( v\times v_{5}\right) & \ 
\end{aligned}\right)\\
&= \frac{1}{2} \left( \left(\textcolor[rgb]{0.29,0.56,0.89}{v_1 \times v_2}\right) + \left(\textcolor[rgb]{0.74,0.06,0.88}{v_2 \times v_3}\right) + \left(\textcolor[rgb]{0.56,0.07,1}{v_5 \times v_1}\right) + \left(\textcolor[rgb]{0.82,0.01,0.11}{v_3 \times v}\right) + \left(\textcolor[rgb]{0.31,0.89,0.76}{v \times v_5}\right) \right)
\end{aligned}
\end{equation*}}
{\small We know from the computation of change in area of a planar boundary that, }
{\small \begin{equation*}
\begin{aligned}
\Eulerconst &= \frac{1}{2}\left(\begin{aligned}
    \left(\cancel{\textcolor[rgb]{0.55,0.34,0.16}{v \times v_1}}\right) +
    \left(v_1 \times v_2\right) +
    \left(\cancel{\textcolor[rgb]{0.96,0.65,0.14}{v_2 \times v}}\right) & + \\
    \left(\cancel{\textcolor[rgb]{0.96,0.65,0.14}{v\times v_2}}\right) +
    \left(v_2 \times v_3\right) +
    \left(v_3 \times v\right) & + \\
    \left(v\times v_5\right) +
    \left(v_5 \times v_1\right) +
    \left(\cancel{\textcolor[rgb]{0.55,0.34,0.16}{v_1 \times v}}\right) &
\end{aligned}\right) \\
 & =\frac{1}{2}\left(\left(\textcolor[rgb]{0.29,0.56,0.89}{v}\textcolor[rgb]{0.29,0.56,0.89}{_{1}}\textcolor[rgb]{0.29,0.56,0.89}{\times v}\textcolor[rgb]{0.29,0.56,0.89}{_{2}}\right) +\left(\textcolor[rgb]{0.74,0.06,0.88}{v}\textcolor[rgb]{0.74,0.06,0.88}{_{2}}\textcolor[rgb]{0.74,0.06,0.88}{\times v}\textcolor[rgb]{0.74,0.06,0.88}{_{3}}\right) +\left(\textcolor[rgb]{0.82,0.01,0.11}{v}\textcolor[rgb]{0.82,0.01,0.11}{_{3}}\textcolor[rgb]{0.82,0.01,0.11}{\times v}\right) \ +\left(\textcolor[rgb]{0.31,0.89,0.76}{v\times v}\textcolor[rgb]{0.31,0.89,0.76}{_{5}}\right) +\left(\textcolor[rgb]{0.56,0.07,1}{v}\textcolor[rgb]{0.56,0.07,1}{_{5}}\textcolor[rgb]{0.56,0.07,1}{\times v}\textcolor[rgb]{0.56,0.07,1}{_{1}}\right)\right)
\end{aligned}
\end{equation*}}
{\small Thus we prove that $\Eulerconst=\Eulerconst '$, meaning the change in area computation is the same for planar and non-planar boundaries.}

\noindent{\small \textbf{Explanation: In volume preservation, volume of a tetrahedron is given by }$\displaystyle V=\frac{1}{6}\begin{vmatrix}
v_{x} & v_{1x}^{t} & v_{2x}^{t} & v_{3x}^{t}\\
v_{y} & v_{1y}^{t} & v_{2y}^{t} & v_{3y}^{t}\\
v_{z} & v_{1z}^{t} & v_{2z}^{t} & v_{3z}^{t}\\
1 & 1 & 1 & 1
\end{vmatrix}$}

{\small To simplify the quantity $\displaystyle V$, we apply the column transformations $\displaystyle C_{2} =C_{2} -C_{1}$, $\displaystyle C_{3} =C_{3} -C_{1}$ and $\displaystyle C_{4} =C_{4} -C_{1}$. This gives us:}

{\small \begin{equation*}
V=\frac{1}{6}\begin{vmatrix}
v_{x} & v_{1x}^{t} -v_{x} & v_{2x}^{t} -v_{x} & v_{3x}^{t} -v_{x}\\
v_{y} & v_{1y}^{t} -v_{y} & v_{2y}^{t} -v_{y} & v_{3y}^{t} -v_{y}\\
v_{z} & v_{1z}^{t} -v_{z} & v_{2z}^{t} -v_{z} & v_{3z}^{t} -v_{z}\\
1 & 0 & 0 & 0
\end{vmatrix}
\end{equation*}}

{\small Evaluating this determinant along the last row reduces it to a $\displaystyle 3\times 3$ determinant:}

{\small \begin{equation*}
V=-\frac{1}{6}\begin{vmatrix}
v_{1x}^{t} -v_{x} & v_{2x}^{t} -v_{x} & v_{3x}^{t} -v_{x}\\
v_{1y}^{t} -v_{y} & v_{2y}^{t} -v_{y} & v_{3y}^{t} -v_{y}\\
v_{1z}^{t} -v_{z} & v_{2z}^{t} -v_{z} & v_{3z}^{t} -v_{z}
\end{vmatrix}
\end{equation*}}

\begin{figure}[!ht]
\centering
   \includegraphics[width=0.65\columnwidth]{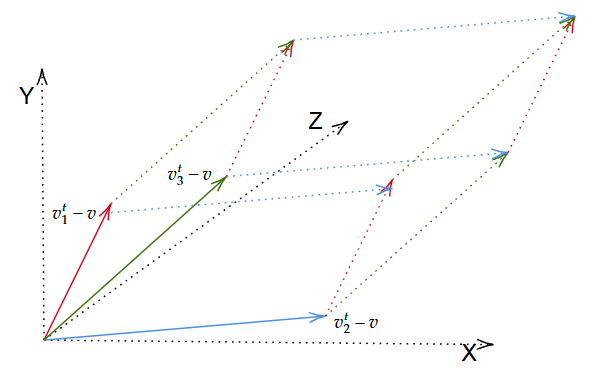}
   \caption{\label{fig:tetrahedron-parallelopiped} Volume of a tetrahedron parallelopiped illustration}
\end{figure}

{\small If we ignore the constant in front, we can note that the determinant of a matrix with basis vectors $\displaystyle C_{1} ,\ C_{2}\text{ and } C_{3}$ gives the signed volume of the parallelepiped they span, as shown in \autoref{fig:tetrahedron-parallelopiped}.}

\begin{figure}[!ht]
\centering
   \includegraphics[width=0.65\columnwidth]{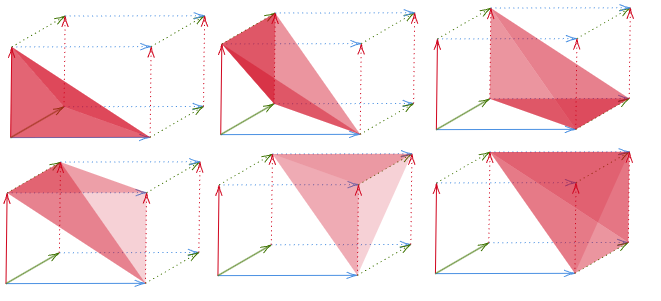}
   \caption{\label{fig:6-equal-tetrahedra} 6 equal tetrahedra from a parallelopiped}
\end{figure}

{\small And, we can show that the parallelepiped geometrically encompasses six tetrahedra of equal volume - see \autoref{fig:6-equal-tetrahedra}.}

{\small Thus, to obtain the volume of a single tetrahedron, we divide the parallelepiped volume by 6.}\newline

\noindent{\small \textbf{Helper functions}}
\begin{lstlisting}
namespace util
{
  template <typename T>
  vector<const T*> GetUnion(const vector<const T*>& vector1, const vector<const T*>& vector2)
  {
    set<const T*> unique_pointers;

    unique_pointers.insert(vector1.begin(), vector1.end());
    unique_pointers.insert(vector2.begin(), vector2.end());

    return vector<const T*>(unique_pointers.begin(), unique_pointers.end());
  }

  bool IsBoundaryEdge(const IMesh* mesh, const IEdge* edge)
  {
    return mesh->GetConnectedFaces(edge).size() < 2;
  }

  vec3 ComputeNormal(const IFace* face)
  {
    auto fv = face->GetVertices();
    vec3 normal = cross(
      fv[1].GetPosition() - fv[0].GetPosition(),
      fv[2].GetPosition() - fv[0].GetPosition()
    );
    return normalize(normal);
  }

  mat3 MakeSkewSymMat(const vec3& v)
  {
    mat3 skew_sym_mat = mat3(0);

    skew_sym_mat[0][1] = -v[2]; skew_sym_mat[0][2] =  v[1];
    skew_sym_mat[1][0] =  v[2]; skew_sym_mat[1][2] = -v[0];
    skew_sym_mat[2][0] = -v[1]; skew_sym_mat[2][1] =  v[0];
  }
}
\end{lstlisting}

\newpage

\section*{Author Contact Information}

\hspace{-2mm}\begin{tabular}{p{0.5\textwidth}p{0.5\textwidth}}
Purva Kulkarni \newline
Independent Researcher \newline
1400 Main St. \newline
Canonsburg, PA 15317 \newline
\href{mailto:purvaskulkarni14@gmail.com}{purvaskulkarni14@gmail.com}
&

Aravind Shankara Narayanan \newline
Independent Researcher \newline
9805 Jake Lane \newline
San Diego, CA 92126 \newline
\href{mailto:aravind.rssn@gmail.com}{aravind.rssn@gmail.com}

\end{tabular}

\end{document}